\documentclass[final,5p,times,twocolumn]{elsarticle}

\usepackage{graphicx}  
\usepackage{amssymb}   
\usepackage{amsmath}
\usepackage{hyperref}
\usepackage{verbatim}
\usepackage{multirow}
\usepackage{float}
\usepackage{subfigure}
\usepackage[modulo]{lineno}
\usepackage{xcolor}
\def\Plus{\texttt{+}}

\journal{Physics Letters B}

\begin{document}
\begin{frontmatter}
\title{Studying Gamow-Teller transitions and the assignment of isomeric and ground states at $N=50$}



\author[1,2,3]{Ali Mollaebrahimi\corref{cor1}}
\ead{Ali.mollaebrahimi@exp2.physik.uni-giessen.de}
\cortext[cor1]{Corresponding author}
\author[2]{Christine Hornung}
\author[2,3]{Timo Dickel}
\author[2]{Daler Amanbayev}
\author[2]{Gabriella Kripko-Koncz}
\author[2,3]{Wolfgang R.\ Pla\ss}
\author[2,3]{Samuel Ayet San Andr\'{e}s}
\author[2,3]{S\"onke Beck}
\author[4]{Andrey Blazhev}
\author[2]{Julian Bergmann}
\author[2,3]{Hans Geissel}
\author[3]{Magdalena G\'orska}
\author[3]{Hubert Grawe\fnref{fn1}}
\fntext[fn1]{Deceased}
\author[2]{Florian Greiner}
\author[3]{Emma Haettner}
\author[1]{Nasser Kalantar-Nayestanaki}
\author[2]{Ivan Miskun}
\author[5]{Fr\'ed\'eric Nowacki}
\author[2,3,6]{Christoph~Scheidenberger}
\author[2,3,7,8]{Soumya Bagchi}
\author[9]{Dimiter L. Balabanski}
\author[10]{Ziga Brencic}
\author[11]{Olga Charviakova}
\author[9]{Paul Constantin}
\author[3]{Masoumeh Dehghan} 
\author[2]{Jens Ebert}
\author[2]{Lizzy Gr\"of}
\author[12]{Oscar Hall}
\author[1]{Muhsin N. Harakeh}
\author[8]{Satbir Kaur}
\author[13,14]{Anu Kankainen}
\author[3]{Ronja Kn\"obel}
\author[2,3]{Daria A. Kostyleva}
\author[15]{Natalia Kurkova}
\author[3]{Natalia Kuzminchuk}
\author[16,17]{Israel Mardor}
\author[9,18]{Dragos Nichita}
\author[2]{Jan-Hendrik Otto}
\author[11]{Zygmunt Patyk}
\author[3]{Stephane Pietri}
\author[3]{Sivaji Purushothaman}
\author[12]{Moritz Pascal Reiter}
\author[2]{Ann-Kathrin~Rink}
\author[3,19]{Heidi Roesch}
\author[9,18]{Anamaria Sp\u{a}taru}
\author[20]{Goran Stanic}
\author[9,18]{Alexandru State}
\author[21]{Yoshiki K. Tanaka}
\author[10]{Matjaz Vencelj}
\author[3]{Helmut Weick}
\author[3]{John S.\ Winfield\fnref{fn2}}
\fntext[fn2]{Deceased}
\author[22]{Michael I.\ Yavor}
\author[3]{Jianwei Zhao}

\affiliation[1]{
organization={Nuclear Energy Group, ESRIG, University of Groningen},
addressline={Zernikelaan 25},
postcode={9747 AA},
city={Groningen},
country={The Netherlands}}

\affiliation[2]{
organization={II. Physikalisches Institut, Justus-Liebig-Universität},
addressline={Heinrich-Buff-Ring 16},
city={Gie{\ss}en},
postcode={35392},
country={Germany}}

\affiliation[3]{
organization={GSI Helmholtzzentrum für Schwerionenforschung GmbH},
addressline={Planckstraße 1},
city={Darmstadt},
postcode={64291},
country={Germany}}

\affiliation[4]{
organization={Institut für Kernphysik, Universität zu Köln},
addressline={Universitätsstraße 16},
city={Köln},
postcode={D-50937},
country={Germany}}            

\affiliation[5]{
organization={Université de Strasbourg, CNRS},
addressline={IPHC UMR},
city={Strasbourg},
postcode={7178, F-67000},
country={France}} 

\affiliation[6]{
organization={Helmholtz Research Academy Hesse for FAIR (HFHF), GSI Helmholtz Center for Heavy Ion Research},
addressline={},
city={Gie{\ss}en},
postcode={35392},
country={Germany}}

\affiliation[7]{
organization={Indian Institute of Technology (Indian School of Mines)},
addressline={Jharkhand},
city={Dhanbad},
postcode={826004},
country={India}}

\affiliation[8]{
organization={Saint Mary's University},
addressline={923 Robie St},
city={Halifax},
postcode={NS B3H 3C3},
country={Canada}}

\affiliation[9]{
organization={Extreme Light Infrastructure-Nuclear Physics (ELI-NP)},
addressline={Strada Reactorului 30},
city={Bucharest-Măgurele},
postcode={077125},
country={Romania}}

\affiliation[10]{
organization={Jozef Stefan Institute},
addressline={Jamova cesta 39},
city={Ljubljana},
postcode={SI-1000 Ljubljana},
country={Slovenia}}

\affiliation[11]{
organization={National Centre for Nuclear Research},
addressline={Pasteur 7},
city={Warszawa},
postcode={02-093},
country={Poland}}

\affiliation[12]{
organization={University of Edinburgh},
addressline={},
city={Edinburgh},
postcode={EH9 3FD},
country={United Kingdom}}

\affiliation[13]{
organization={University of Jyväskylä},
addressline={Seminaarinkatu 15},
city={Jyväskylä},
postcode={40014},
country={Finland}}

\affiliation[14]{
organization={Helsinki Institute of Physics},
addressline={},
city={Helsinki},
postcode={00014},
country={Finland}}

\affiliation[15]{
organization={Flerov Laboratory of Nuclear Reactions},
addressline={JINR},
city={Dubna},
postcode={141980},
country={Russia}}

\affiliation[16]{
organization={Tel Aviv University},
addressline={},
city={Tel Aviv},
postcode={6997801},
country={Israel}}

\affiliation[17]{
organization={Soreq Nuclear Research Center},
addressline={},
city={Yavne},
postcode={81800},
country={Israel}}

\affiliation[18]{
organization={Doctoral School in Engineering and Applications of Lasers and Accelerators, University Polytechnica of Bucharest},
addressline={},
city={Bucharest},
postcode={060811},
country={Romania}}

\affiliation[19]{
organization={Technische Universität Darmstadt},
addressline={Karolinenpl. 5},
city={Darmstadt},
postcode={D-64289},
country={Germany}}

\affiliation[20]{
organization={Johannes Gutenberg-Universität Mainz},
addressline={},
city={Mainz},
postcode={55099},
country={Germany}}

\affiliation[21]{
organization={High Energy Nuclear Physics Laboratory, RIKEN},
addressline={2-1 Hirosawa, Wako},
city={Saitama},
postcode={351-0198},
country={Japan}}

\affiliation[22]{
organization={Institute for Analytical Instrumentation, RAS},
addressline={},
city={Petersburg},
postcode={190103},
country={Russia}}

\begin{abstract}
Direct mass measurements of neutron-deficient nuclides around the $N=50$ shell closure below $^{100}$Sn  were performed at the FRS Ion Catcher (FRS-IC) at GSI, Germany. The nuclei were produced by projectile fragmentation of $^{124}$Xe, separated in the fragment separator FRS and delivered to the FRS-IC. The masses of 14 ground states and two isomers were measured with relative mass uncertainties down to $1\times 10^{-7}$ using the multiple-reflection time-of-flight mass spectrometer of the FRS-IC, including the first direct mass measurements of $^{98}$Cd and $^{97}$Rh. A new $Q_\mathrm{EC} = 5437\pm67$ keV was obtained for $^{98}$Cd, resulting in a summed Gamow-Teller (GT) strength for the five observed transitions ($0^+\longrightarrow1^+$) as $B(\text{GT})=2.94^{+0.32}_{-0.28}$. Investigation of this result in state-of-the-art shell model approaches sheds light into a better understanding of the GT transitions in even-even isotones at $N=50$. The excitation energy of the long-lived isomeric state in $^{94}$Rh was determined for the first time to be $293\pm 21$~keV. This, together with the shell model calculations, allows the level ordering in $^{94}$Rh to be understood.  
\end{abstract}


\end{frontmatter}


\section{Introduction} \label{sc_intro}
	The mass of a nucleus reflects its total binding energy and can be measured directly by high-precision mass spectrometry. The mass is one of the most fundamental properties in the study of nuclear structure \cite{Scheidenberger2005}. Isotopes in the medium-heavy and neutron-deficient region near the $N=50$ and $N=Z$ lines are of special interest for nuclear structure studies as this region of the nuclear chart exhibits numerous unique phenomena \cite{wigner,wigner2,94Ag-nature,80Zr,Haettner2011b}. Of these isotopes, the heaviest doubly-magic $N=Z$ nucleus, $^{100}$Sn, is the most desirable to study and attracts considerable attention \cite{Faestermann2013,physics4010024}. Independent measurements of the properties of $^{100}$Sn, such as the $Q_\mathrm{EC}$ value \cite{GT, Lubosss} or the production cross-section \cite{18nbarn} do not agree, which called for additional investigations \cite{nature-100In}. 
	Associated with the contradicting experimental results for the $Q_\mathrm{EC}$ are also large discrepancies in the Gamow-Teller strength $B(\text{GT})$. In general, strong resonances in Gamow-Teller (GT) transitions are observed for the isotopes decaying by $\beta^+$ or Electron Capture (EC) processes. The resonance is known to occur as a result of the interaction between the nucleons in the configuration of $1g_{9/2}$ and $1g_{7/2}$ orbitals near the $N=Z=50$ shell closure \cite{odd-even,Batist2010,KARNY19983,PhysRevC.60.024315} and the accessible Gamow-Teller strength distribution within the large $Q_\mathrm{EC}$ values close to the proton drip-line.
Necessary input parameters for the determination of the $B(\text{GT})$ are the $Q_\mathrm{EC}$ values, half-lives, decay branching ratios and the level scheme of daughter nuclei. The most precise determination of $Q_\mathrm{EC}$ value is obtained by direct mass measurements of the mother and daughter nuclei. However, the production of exotic isotopes close to the proton drip-line and in the vicinity of the $N=Z$ line is challenging due to the very low production cross sections. As a result, their mass has often been measured indirectly with poor mass accuracy and not for all relevant isotopes so far.

By measuring the masses of nuclei, one can also study the appearance of excited metastable states, known as nuclear isomers. The properties of nuclear isomers \cite{Walker2020,Walker2010,Dracoulis2016} are also of great importance for understanding the structure of nuclei. Nuclear-shape deformations and excitation of nucleons in their shell levels can result in a major spin change compared to the corresponding ground state. 
Isomeric states can be investigated by high-precision mass spectrometry, yielding direct proof of their existence and determination of their excitation energy relative to the ground state. During the last two decades, high-resolution direct mass measurements of isomers were mainly performed with storage rings \cite{isomer_storage_ring,Franzke2008} and Penning traps \cite{PhysRevLett.100.132501,Blaum2006}. The recent improvement of the MR-TOF-MS (Multiple-Reflection Time-Of-Flight Mass Spectrometer) technique with high resolving power \cite{Mardor-PRC} has facilitated the discovery and study of low-lying isomeric states \cite{Christine,SoenkePRL}. The FRS Ion Catcher (FRS-IC) at the fragment separator \cite{Geissel1992} at GSI offers unique conditions for production and high-precision measurements of the ground-state and isomeric-state masses of exotic nuclei \cite{Dickel2015b,Plass2008,Plass2013}. 

The nuclear structure at $N=50$ was studied experimentally by studying 14 isotopes and 2 isomers in that region of the nuclear chart, and the results where used to benchmark state-of-the-art large scale shell-model calculations and thus provide an improved basis for the predictions of the nuclear structure of $^{100}$Sn.
	
\section{Experiment} \label{sc_exp}

    

    The FRS Ion Catcher (FRS-IC) \cite{Plass2013} is an experimental setup installed at the final focal plane of the fragment separator FRS \cite{Geissel1992b} at GSI. The FRS in combination with the FRS-IC enables experiments with trapped exotic nuclei. The FRS-IC consists of three main parts: (i) the gas-filled Cryogenic Stopping Cell (CSC) \cite{Ranjan2011,Purushothaman2013,Ranjan2015} for slowing-down and thermalization of the exotic nuclei produced at relativistic energies, (ii) a radio-frequency quadrupole (RFQ) beamline \cite{Plass2008,Reiter2015,Miskun2015,Haettner2018} for mass-selective transport and differential pumping, (iii) the MR-TOF-MS \cite{Dickel2015b,Plass2008} for performing direct mass measurements, which entails a unique combination of performance parameters - fast ($\sim$ms), accurate, broadband and non-scanning \cite{Dickel2015b} operation. The mass measurement is done by injecting ions into the isochronous Time-Of-Flight (TOF) analyzer and confining them for a given number of turns between two electrostatic mirrors. The ions with different mass-to-charge ratio ($m/q$), obtain different velocities and are spatially separated in their path-line in the TOF analyzer. The mass-range selector (MRS) \cite{Dickel2015b} at the center of the analyzer can also cut a certain range of masses by applying deflecting pulses while ions pass through it. The ions are then ejected from the TOF analyzer by opening the exit mirror toward the detector for recording the TOF for the different $m/q$ species. 
    
    In the experiments reported here, the exotic nuclei and their isomeric states were produced via projectile fragmentation in two experiments. 
    Experiment I was performed with an 800~MeV/u $^{124}$Xe projectile beam with an intensity of up to $1.5\cdot10^9$ ions per spill, a typical spill length of 3 s and a repetition rate of 0.2 Hz on a beryllium production target with an areal density of 8045\,mg/cm$^2$. The FRS was centered on $^{98}$Cd.  The mono-energetic degrader at the central focal plane had an areal density of 737~mg/cm$^2$. The fragments were then stopped and thermalized in the cryogenic stopping cell of the FRS-IC operating with helium at 85$\pm$1\,K and 110$\pm$5\,mbar corresponding to an areal gas density of 6.57$\pm$0.34~mg/cm$^2$, and with a mean ion extraction time of about 80\,ms from the CSC. 
    The ions were then transported to MR-TOF-MS for the high-precision mass measurements.
    Experiment II was performed with a 600~MeV/u $^{124}$Xe projectile beam with an intensity of up to $1\cdot10^9$ ions per spill, with a typical spill length of 500~ms and a repetition rate of 0.25 Hz on a beryllium production target with an areal density of 1622~mg/cm$^2$. The centered fragment in the FRS was $^{94}$Rh. The mono-energetic degrader at the central focal plane had an areal density of 737~mg/cm$^2$. The CSC operated at a pressure of 75$\pm$5~mbar and a temperature of 82$\pm$1~K, corresponding to an areal density of 4.64$\pm$0.31~mg/cm$^2$ helium. This results in a mean ion extraction time of about 200\,ms from CSC. In both experiments, a mass resolving power of 400,000 to 500,000 was achieved.

     The development and validation of the MR-TOF-MS data analysis procedure is presented in a separate publication \cite{Ayet2019}. To convert time-of-flight into mass calibrant ions are used; all (see Tab. 1) have have negligible small mass uncertainties. Mass values, their uncertainties and rates of the measured isotopes are obtained after fitting the mass peaks to a Hyper Exponentially-Modified Gaussian (Hyper-EMG) function \cite{Purushothaman2017}. The procedure allows accurate mass determination even for the most challenging conditions, including very low numbers of events and overlapping mass peaks. Unresolved isomeric states are reflected as an additional uncertainty contribution to the evaluation of the ground-state mass (AME, appendix B4 \cite{AME2020}). In the mass measurements reported in Ref. \cite{Mardor-PRC}, a total relative uncertainty down to $1.7\cdot 10^{-8}$ was achieved. 
    
\section{Results}

  The mass measurements are summarized in Table \ref{tab:94Rh_results}. The Birge ratio \cite{BR} is calculated for the isotopes with known direct mass measurements to be 0.86, thus showing the excellent agreement with literature values. 
For $^{97}$Ag, the isomer-to-ground-state ratio of $0.079\pm0.019$ was obtained and is in agreement with Ref. \cite{Christine}. As the mass accuracy reported here is similar to the one reported in \cite{Christine} we calculate the weighted average of the mass excess value to $-70914 \pm10$~keV 

\begin{figure} 
    	\centering
    	\includegraphics*[width=8.6cm]{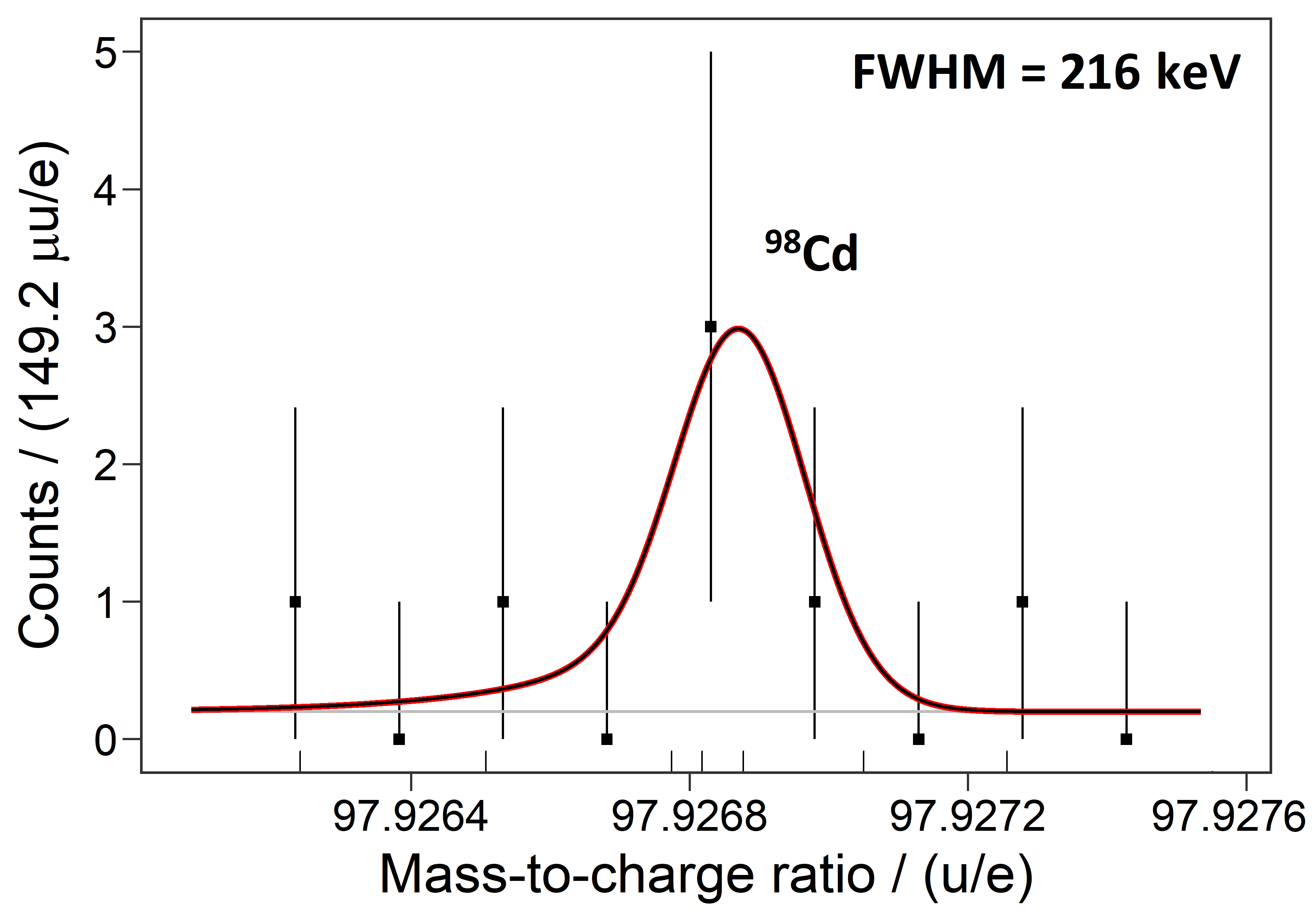}
    	
    	\includegraphics*[width=8.6cm]{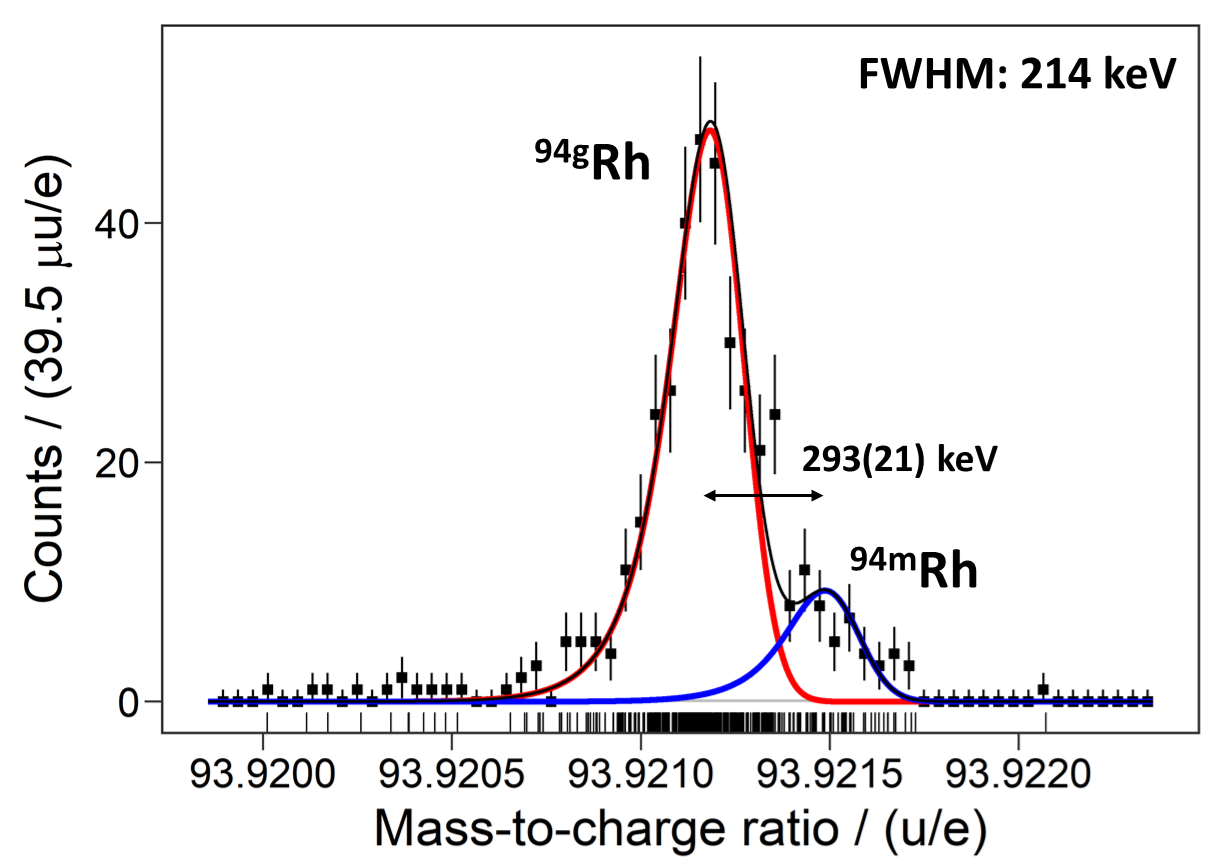} 
    	\caption{Mass spectrum of $^{98}$Cd ions (top panel) and $^{94}$Rh ions (bottom panel). For $^{98}$Cd the data were fitted with a Hyper-EMG function with one exponential tail on each side. For $^{94}$Rh the ground state (red curve) and isomeric state (blue curve) were fitted with a double Hyper-EMG with one exponential tail on the left side. An isomer-to-ground state ratio of 0.19(4) was determined. The histogram of the measured spectrum (black data points) is only drawn to guide the eye; the data analysis was based on the unbinned data (rug graph) below the mass spectrum using the weighted Maximum Likelihood procedure described in Ref. \cite{Ayet2018}. }
    	\label{fig:94Rh}
    \end{figure}

  The excitation energy of the isomeric state in $^{94g,m}$Rh is measured for the first time and is determined to be $293\pm 21$~keV. The measured mass distribution of $^{94g,m}$Rh ions and the corresponding fitted curves are shown in Figure~\ref{fig:94Rh}, where two measurements are combined and weighted by their errors. 
    For the $^{94}$Rh nucleus, two states with half-lives of the order of a minute are reported in literature. The relative order of the two states and their assigned spins is currently under discussion. In NUBASE 2020 \cite{NUBASE2020}, the ground state is assigned to the $\left( 4\right)^{\Plus}$ state with a half-life of 70.6~s. For the isomeric state, a spin-parity of $\left( 8\right)^{\Plus}$ and a half-life of 25.8~s with only an extrapolated value for excitation energy (\#300 keV) has been reported. The spin-parity assignment of the ground state is based on a $\beta$~end-point measurement of the $\left( 4\right)^{\Plus}$ state, which was reported originally as a $\left( 3\right)^{\Plus}$ state, with a half-life of 70.6~s \cite{Oxorn1980}. In this reference, a $Q_\mathrm{EC}$-value of 10.0$\pm$0.4~MeV for $^{94}$Rh decaying to $^{94}$Ru was reported and it was assigned to the isomeric state of $^{94m}$Rh. However, shell model calculations \cite{Herndl1997} disagree with this assignment, predicting the state with 70.6~s half-life to be an excited state at 145~keV with spin-parity $\left( 4\right)^{\Plus}$ and the $\left( 8\right)^{\Plus}$ state to be the ground state with 25.8~s half-life.
    
    The ground state of $^{97}$Rh is measured for the first time in a direct mass measurement. The previous mass evaluation is from the $\beta$-end-point energy $Q_\mathrm{EC}=3520\,\pm$\,40\,keV. This value is a weighted average of two different measurements \cite{97Rh-indirect1,97Rh-indirect2}. The ground state of $^{97}$Rh and the low-lying isomeric state $^{97m}$Rh are fitted by using a double Hyper-EMG function with a fixed distance between the ground state and the isomeric state of 258.76\,$\pm$\,0.18\,keV known from gamma spectroscopy \cite{97Rh-gamma}. This allows the determination of the ground-state mass even though the isomeric state is not fully resolved. An isomer-to-ground state ratio of $0.064\pm0.043$ was determined. Our measurement improves the uncertainty by more than a factor 2  and shows that the nucleus is 50 keV less bound compared to the previous indirect measurements. 
    
The mass of ground-state of $^{98}$Cd is measured directly for the first time. The previous indirect measurements, which are the basis of the mass value reported in the AME2020 \cite{AME2020}, are from the $\beta$-end-point energy measurement with $Q_\mathrm{EC}=5430\,\pm$\,40\,keV published in a conference proceeding \cite{AIP-proceeding} and in a non-peer-reviewed laboratory report \cite{98Cd-idirect-GSI} and $Q_\mathrm{EC}=5330\,\pm$\,140\,keV evaluated based on a semi-empirical method \cite{98Cd-indirect}. 
The high sensitivity and reliability of the MR-TOF-MS setup provides the possibility of a mass measurement with a very-low number of identified ions. This has been verified in previous experiments using the FRS-IC \cite{Christine, Mardor-PRC}. A further benchmark is the high mass-accuracy achieved with only 5-6 counts ($^{97}$Pd and $^{100}$Pd) in this paper. $^{98}$Cd (7 counts) was analyzed using the same technique, see (Figure~\ref{fig:94Rh}). An expected background of $1.7\pm0.2$ counts under the peak is determined from the measurement. The probability of having 7 background events based on the expected background is only 0.15\%. The possibility that an unexpected contamination peak lies beneath the $^{98}$Cd is included in the uncertainty of the mass value. From the particle identification of the FRS and the known efficiencies of the setup we expect 4 events of $^{98}$Cd, thus in very good agreement with the measured rate. A mass excess value of $-67633\,\pm$\,60\,keV was determined for the ground state of $^{98}$Cd. Note that the production of $^{98}$Cd by projectile fragmentation of $^{124}$Xe has a cross section of 18 nb \cite{18nbarn} only, which further illustrates the high sensitivity of the FRS-IC.

    \begin{table*}[t] 
    	\centering 
        \begin{tabular}{p{1cm}ccccccc} \hline\hline
        Nuclide  &Calibrant & Number of   & ME/EX$_\mathrm{FRS-IC}$ & ME/EX$_\mathrm{AME20}$ & Difference & Number & Exp.\\
        & &Isochronous Turn & / keV & / keV & / keV & of events & \\ 
         \hline \hline
         
        $^{94}$Ru  &$^{12}$C$_3^{19}$F$_3$&625 ($A=94$)& $-82556$ $\pm$ 34 & $-82584$ $\pm$ 3 & 28 $\pm$ 34 & 69 &(II)\\ \hline
        
    	$^{94}$Rh$^*$ & $^{12}$C$_2^{16}$O$^{19}$F$_3$/$^{12}$C$_3^{19}$F$_3$ &625 ($A=94$)& $-72885$ $\pm$ 20 & $-72908$ $\pm$ 3 & 23 $\pm$ 20 & 338 &(II)\\ \hline
    	
    	$^{94m}$Rh$^*$ & $^{12}$C$_2^{16}$O$^{19}$F$_3$/$^{12}$C$_3^{19}$F$_3$&625 ($A=94$)& 293 $\pm$ 21 & - & - &64 &(II) \\ \hline
    	
    	$^{96}$Ru  &$^{96}$Pd &800 ($A=97$)& $-86075$ $\pm$ 28 & $-86080$ $\pm$ 0.17 & 5 $\pm$ 28 & 15 &(I)\\ \hline
    	
    	$^{96}$Rh  &$^{96}$Pd&800 ($A=97$)& $-79655$ $\pm$ 40 & $-79688$ $\pm$ 10 & 33 $\pm$ 41 & 10 &(I)\\ \hline
    	
    	$^{97}$Rh  &$^{96}$Pd&800 ($A=97$)& $-82550$ $\pm$ 19 & $-82600$ $\pm$ 40 $^{\ddagger}$ & 50 $\pm$ 44 & 77 &(I)\\ \hline
        
        $^{97m}$Rh  &$^{96}$Pd&800 ($A=97$)& fixed & 258.76 $\pm$ 0.18 & - & 4 &(I)\\ \hline
        
    	$^{97}$Pd  &$^{96}$Pd &800 ($A=97$)& $-77824$ $\pm$ 59 & $-77806$ $\pm$ 5 & $-18$ $\pm$ 59 & 6 &(I)\\ \hline
    	
    	$^{97}$Ag  &$^{96}$Pd &800 ($A=97$)& $-70940$ $\pm$ 19 & $-70904$ $\pm$ 12 & $-36$ $\pm$ 22 & 54 &(I)\\ \hline
    	
    	$^{97m}$Ag  &$^{96}$Pd &800 ($A=97$)& 608 $\pm$ 73 & 620 $\pm$ 40 & $-12\pm$ 48 & 4 &(I)\\ \hline
    	
    	$^{98}$Pd  &$^{96}$Pd &800 ($A=97$)& $-81326$ $\pm$ 21 & $-81321$ $\pm$ 5 & $-5$ $\pm$ 22 & 102 &(I)\\ \hline
    	
    	$^{98}$Cd &  $^{96}$Pd &800 ($A=97$)& $-67633$ $\pm$ 60 & $-67640$ $\pm$ 50 $^{\dagger}$ & 7 $\pm$ 78 & 7 &(I)\\ \hline
    	
    	$^{99}$Ag  &$^{12}$C$_2^{19}$F$_4$ &800 ($A=99$)& $-76709$ $\pm$ 33 & $-76712$ $\pm$ 6 & 3 $\pm$ 33 & 53 &(I)\\ \hline
    	
    	$^{100}$Pd  &$^{12}$C$_2^{19}$F$_4$&850 ($A=100$)& $-85202$ $\pm$ 60 & $-85213$ $\pm$ 18 & 11 $\pm$ 63 & 5 &(I)\\ \hline
    	
    	$^{100}$Ag  &$^{12}$C$_2^{19}$F$_4$&800 ($A=99$)& $-78095$ $\pm$ 31 & $-78138$ $\pm$ 5 & 43 $\pm$ 32 & 12 &(I)\\ \hline
    	
    	$^{100}$Cd  &$^{12}$C$_2^{19}$F$_4$&800,850 ($A=99,100$)& $-74194$ $\pm$ 12 & $-74195$ $\pm$ 2 & 0.7 $\pm$ 12 & 74 &(I)\\ \hline
    	
    	$^{101}$In  &$^{12}$C$_2^{19}$F$_4$&850 ($A=100$)& $-68559$ $\pm$ 25 & $-68545$ $\pm$ 12 & $-14$ $\pm$ 28 & 17 &(I)\\ \hline
    	
    	\hline \hline
    	\end{tabular}
    	\caption{Measured mass excess (ME) values for ground states and the excitation energies of isomers (EX). $^{\dagger}$ Value reported in AME2020 \cite{AME2020} based on an indirect measurement (private communication) \cite{98Cd-idirect-GSI} and semi-empirical method \cite{98Cd-indirect}. $^{\ddagger}$ Indirect measurement from beta end-point measurements. $^*$ Assignment to ground and isomeric  states previously uncertain. For the nuclides with previously known masses a Birge ratio of 0.86 was obtained.}
    	\label{tab:94Rh_results}
    \end{table*}

 \section{Assigned Level scheme for the nuclide $^{94}\mathrm{Rh}$}  
    
    From our mass measurement, it was possible to directly measure the excitation energy of $293\pm 21$~keV for the isomeric state of $^{94m}$Rh for the first time, a similar excitation energy of $270.09\pm0.08$~keV as for the $\left( 4\right)^{\Plus}$ isomer in the neighboring $^{92}$Tc isotope. This is already a strong indication that both isotopes should have a similar level scheme for the first excited states; $\left(8\right)^{\Plus}$ ground state and $\left( 4\right)^{\Plus}$ isomeric state \cite{NUBASE2020}. From other isotonic chains in this region of the chart of nuclei, one would expect an increase in excitation energy with increasing number of protons. This expected effect is in agreement with the excitation energy of $^{94}$Rh measured here compared to the excitation energy in $^{92}$Tc.
    
    Previously, $^{94}$Rh was investigated in Penning-trap measurements at JYFLTRAP \cite{Weber2008}. There, only one state was measured in the fusion reaction $^{40}$Ca$\Plus ^{58}$Ni. This reaction was also studied at SHIP at GSI \cite{Kurcewicz1982}. In this measurement, the grow-in behaviour of the 25.8~s and the 70.6~s states were compared. The 25.8~s (high spin) state is produced directly in the reaction, while the 70.6~s state (lower spin) showed a feeding from the $\beta$ decay of $^{94}$Pd. 
    Based on the results from Kurcewicz et al. \cite{Kurcewicz1982} and employing an additional study of the states in $^{94}$Ru, Weber et al. \cite{Weber2008} assigned the mass measured at JYFLTRAP to the $\left( 8\right)^{\Plus}$ state, following the argument that fusion-evaporation reactions generally tend to produce higher spin states. In addition to the ME value of the $\left( 8\right)^{\Plus}$ state in $^{94}$Rh, the ground state masses of $^{94}$Pd and $^{94}$Ru were measured at JYFLTRAP and SHIPTRAP \cite{Weber2008}. With the subsequent measurement of these three isotopes, it was possible to determine $Q_\mathrm{EC}$-values of the $\beta$-decays of $^{94}$Pd and of the $\left( 8\right)^{\Plus}$ state of $^{94}$Rh. This has yielded $Q_\mathrm{EC}$-values of 6809.6$\pm$6.3~keV and 9673.1$\pm$5.9~keV, respectively. The $Q_\mathrm{EC}$-values of the $\beta$ decays of $^{94}$Pd and of the 70.6~s state of $^{94}$Rh, which was assigned to a spin-parity of $\left( 4\right)^{\Plus}$, were obtained via the total $\gamma$-ray absorption techniques \cite{Batist2006}. They measured $Q_\mathrm{EC}$-values of 6700$\pm$320~keV and 9750$\pm$320~keV, respectively. The ME value of $^{94}$Rh was investigated by the Canadian Penning Trap (CPT) mass spectrometer \cite{Clark2005,Fallis2011} as well. In this measurement, only one state was observed and they followed the same arguments as Weber et al. to assign the measured value to the $\left( 8\right)^{\Plus}$ in $^{94}$Rh. 

\begin{figure} 
    \centering 
    \includegraphics*[width=8.6cm]{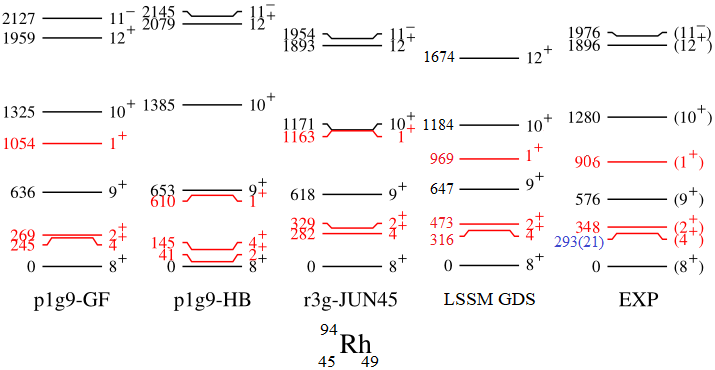} 
    \caption[width=0.7\linewidth]{Experimental level scheme of $^{94}$Rh versus shell model calculations for various interactions and model spaces. The suggested assignment of the spins of the ground and isomeric states is based on previous Penning trap measurements of the long-lived state and  the direct measurement of both states and thus the excitation energy in this work. The red color shows the entangled states affected by the direct measurement of isomer. The further level information is taken from the ENSDF, published in Ref. \cite{Abriola2006}.}
    \label{fig:94Rh_levels}
    \end{figure}

    The ME value of the present experiment for the ground state of $^{94}$Rh is 23$\pm$20~keV higher than the AME2020 value. For the isomeric state, the literature data are very controversial. The directly-measured excitation energy with the MR-TOF-MS in combination with the spin-parity assignment above disentangles the level scheme of $^{94}$Rh and fixes all levels on an absolute scale, which have only been measured relative to the isomeric state. The resulting level scheme is shown in Figure~\ref{fig:94Rh_levels}. The additional information on the level scheme are taken from the Evaluated Nuclear Structure Data File (\emph{ENSDF}), published in Ref. \cite{Abriola2006}. 
    
    Calculations in several different shell model approaches were performed. The $\pi \nu (p_{1/2}, g_{9/2})$ model space and a $^{76}$Sr core were used with an empirically fitted isospin-asymmetric interaction \cite{Gross1976} (GF) and an isospin-symmetric empirical fit (HB) \cite{Herndl1997}. The latter served as the basis for a comprehensive binding energy extrapolation in the triangle $^{76}$Sr, $^{88}$Sr and $^{100}$Sn relative to $^{88}$Sr \cite{Herndl1997}. Furthermore, Large Scale Shell Model (LSSM) calculations have been performed in the full $\pi\nu$ $r3g$ model space between $^{56}$Ni and $^{100}$Sn using the JUN45 interaction \cite{PhysRevC.80.064323} and in the full $n=4$ harmonic oscillator  $sdg$ shell. This latter valence space has the advantage of incorporating $Z,N=50$ core excitations and 
    contains full $0\hbar\omega$ correlations for E2 and Gamow-Teller operators. The corresponding  hamiltonian  is the SDGN effective interaction \cite{Blazhev:2004gm,Siciliano:2019qhw,RISING:2021ait,GT} which provides excellent spectroscopy in the $A=90-120$ region.
    The largest diagonalisations achieved in this study were for the Yrast band of $^{94}$Rh with $15\cdot10^9$ basis states dimension.\\
    The results yield a good agreement with experiment except for the relative isomer position in the HB approach. This, however, should not be an argument for the existence of isospin-asymmetry in the interaction, as JUN45 is an isospin symmetric interaction and reproduces the isomer energy very well. 
    The deviations for the newly established energy of the 1$^+$ state in $^{94}$Rh as observed in Gamow-Teller (GT) decay of $^{94}$Pd are due to their content of core-excited configurations. This is demonstrated in the calculation including sdg model space where the agreement is remarkable.
 
\section{Systematic studies of nuclides at the $N=50$ shell closure}

\begin{figure} 
    	\centering
    	\includegraphics[width=8.6cm]{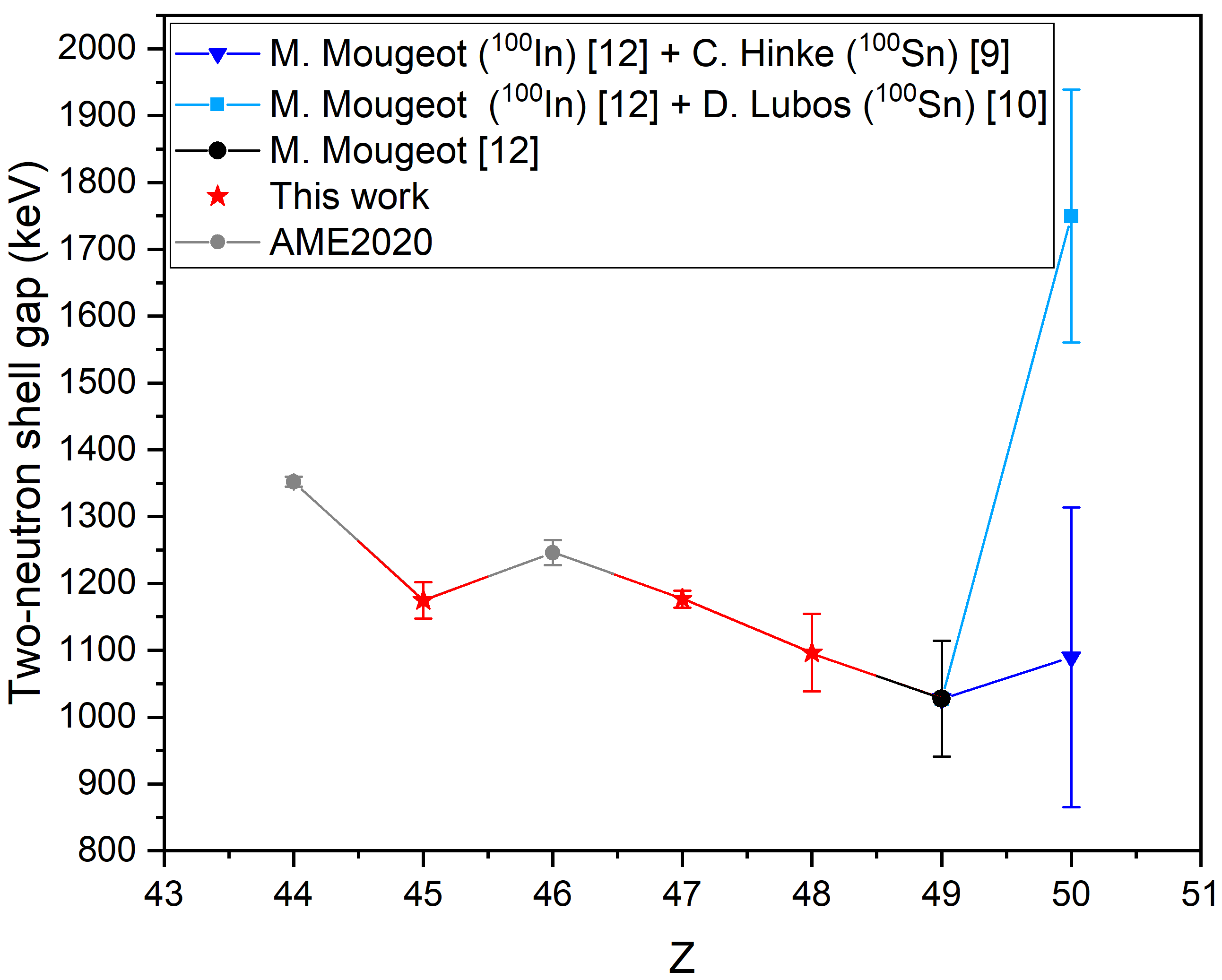} 
    	\caption{The shifted two-neutron shell gap at $N=50$. The gray circles are the literature values \cite{AME2020}, the dark blue triangle and light blue square are from Ref. \cite{nature-100In} rejecting the recent $Q_\mathrm{EC}$ measured \cite{Lubosss} due to a sudden unexpected jump in the shifted two-neutron shell gap. The red stars show $\Delta_{2n}$ influenced by the improved mass values of $^{97}$Ag (weighted average of this work and \cite{Christine}), $^{97}$Rh and $^{98}$Cd reported in this work.}
    	\label{shell-gap}
    \end{figure}

The nuclear structure and shell evolution around the $N=50$ shell closure can be studied by direct mass measurements. Two-neutron separation energy and shell gap are among the observables which can be derived from these mass measurements.
The shifted two-neutron shell gap (shifted for $N+2$ due to lack of experimental data for $N=48$) is described as $\Delta_{2n} (Z,N+2) = M(Z,N)-2 \cdot M(Z,N+2) + M(Z,N+4)$ \cite{nature-100In}. This observable is also the basis for the argument in a recent article by M. Mougeot et al. \cite{nature-100In} about the correct $Q_\mathrm{EC}$ value and the corresponding binding energy reported for the doubly-magic $^{100}$Sn. Currently there exist two contradictory measurements \cite{GT, Lubosss}. Figure~\ref{shell-gap} shows $\Delta_{2n}$ for $N=50$. 
The sudden jump based on the $Q_\mathrm{EC}$ measured in Ref. \cite{Lubosss} was unfavored in comparison to the theoretical ab-initio calculations and the behaviour of $\Delta_{2n}$ for $N=28$. The red stars show $\Delta_{2n}$ influenced by the improved mass values of $^{97}$Ag (weighted average of this work and \cite{Christine}), $^{97}$Rh and $^{98}$Cd reported in this work. Now all $N=50$ isotones are measured directly, besides $^{100}$Sn, with high precision. These results confirm the general trend in the $N=50$ isotones and thus further support the conclusion made in \cite{nature-100In} about favouring the $Q_\mathrm{EC}$ reported in Ref. \cite{Lubosss}.

\section{Gamow-Teller transition's strength}   
The Gamow-Teller transition in beta decay refers to a configuration in which the spins of the emitted electron and antineutrino (or positron and neutrino) are parallel, coupling to total spin $S=1$, leading to an angular momentum change $\Delta J=0,\pm 1$ between the initial and final states of the nucleus. The transition probability or strength of the decay strongly depends on the underlying shell structure and it is usually distributed among several states. The strength of the beta-decay is usually referred to for direct comparison of the transition probability in different nuclei, independent of energy and atomic number $Z$ in mother nuclei, and can be described as follows for a single-state transition \cite{Lubosss}: 
\begin{equation*} \small
\label{GT-equation}
B(\text{GT})=\dfrac{2\pi^3\hbar^7 \ln(2)}{m_e^2c^4G_F^2V_{ud}^2 (G_A/G_V)^2 f t_{1/2}} = \dfrac{3885 \pm14\,s}{f(Z,\epsilon_0)t_{1/2}}
\end{equation*}
where $f(z,\epsilon_0)t_{1/2}$ value is the comparative half-life calculated by knowing the decay energy $Q_\mathrm{EC}$, half-life $t_{1/2}$ and the decay scheme. The $f(z,\epsilon_0)t_{1/2}$ value is proportional to the fifth power of $Q_\mathrm{EC}$, thus very sensitive in $B(\text{GT})$ determinations. $G_A/G_V$ is the weak coupling constant. $G_F$ and $V_{ud}$ are the Fermi coupling constant and the CKM matrix element, respectively. $c$ is speed of light and $m_e$ is electron mass.


\begin{figure} [ht] 
    	\centering
    	\includegraphics[width=8.6cm]{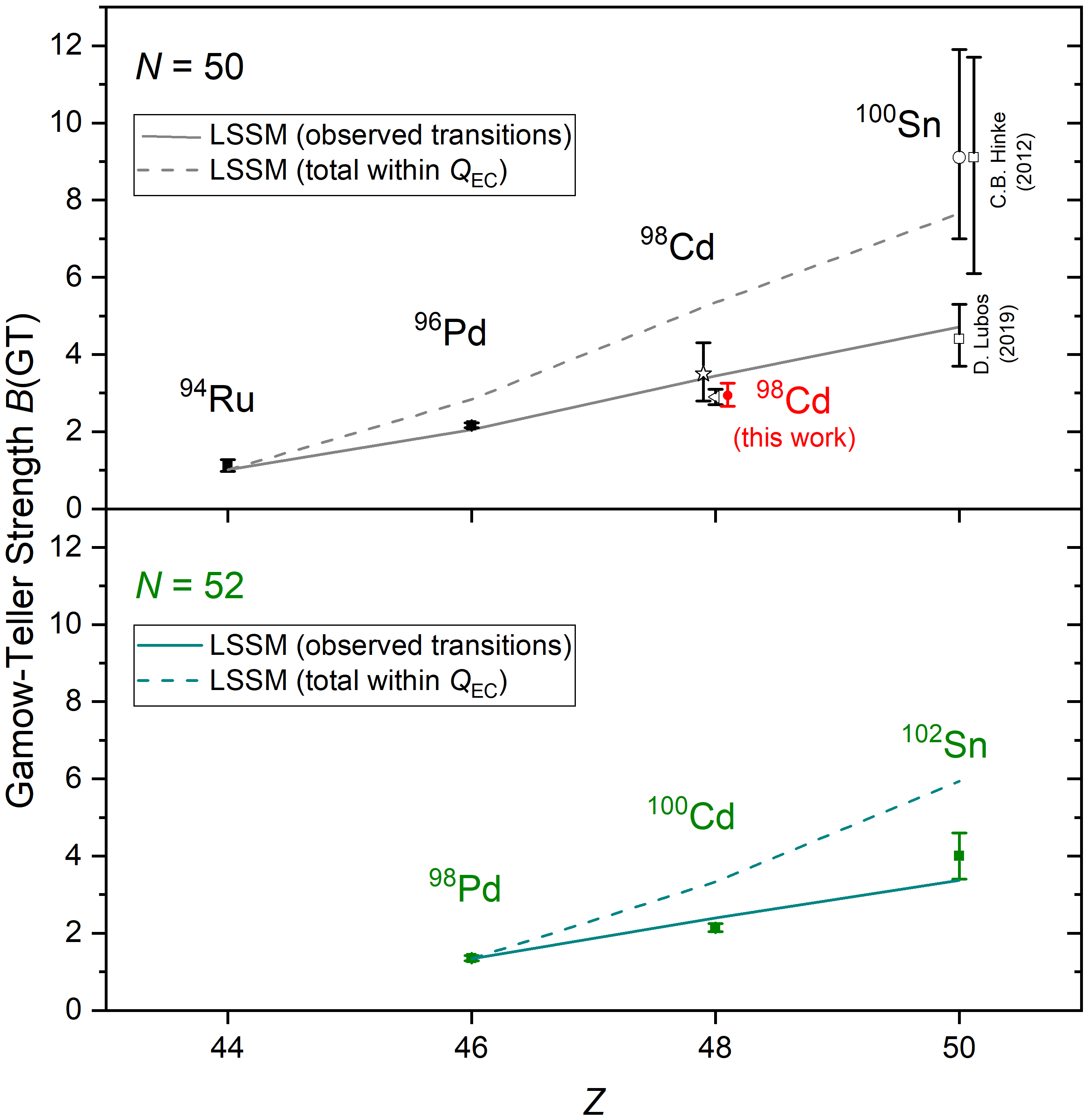} 
    	\caption{The Gamow-Teller strength $B(\text{GT})$ for ($0^+\longrightarrow1^+$) transitions for even-even $N=50$ isotones (top panel) and $N=52$ isotones (bottom panel). The $B(\text{GT})$ values are calculated based on the latest information from the nuclear data sheets. The black star at $Z=48$ shows the previously-reported value for $^{98}$Cd as in Ref. \cite{98Cd-indirect}. The black triangle shows the value reported in Ref. \cite{AIP-proceeding}. The red circle shows the new $B(\text{GT})$ value calculated for $^{98}$Cd based on the first direct mass measurement (this work) and all the recent information. The non-filled black squares shows the $B(\text{GT})$ values for $^{100}$Sn reported in Refs. \cite{GT} and \cite{Lubosss}; the non-filled circle shows the new updated value \cite{GT} considering the improved half-life uncertainty \cite{NUBASE2020}. The shell model calculations summing the GT strength for the experimentally observed decaying states (solid line) or for the total strength contained in the Q$_{EC}$ window (dashed line).}
    	\label{BGT}
    \end{figure}

The experimentally observed GT strengths accessible within the respective $Q_\mathrm{EC}$ windows have shown to be systematically lower than the predicted ones by the nuclear shell model (see for example Ref.~\cite{TOWNER1987263}). A hindrance factor resulting from dividing the theoretical by the experimental $B$(GT) values is calculated and often discussed as having two origins. The first one due to the calculation being performed in a limited (truncated) model space as compared to a calculation in a full (non-truncated) model space, containing all orbitals within a major oscillator shell. The second is due to higher-order correlations that have very similar average values for the $p$-, $sd$-, $fp$-shells (see Refs. \cite{PhysRevC.28.1343,PhysRevC.47.163,PhysRevC.53.R2602}). This higher-order ``global'' suppression of the expected beta-decay rates was often simulated by an effective coupling constant quenched by a factor of about 0.75 \cite{TOWNER1987263, Batist2010}. Recently, this quenching factor was explained using chiral effective field theory combined with two-body currents although small corrections are still possible due to neglected higher-order contributions to currents and Hamiltonians in that approach~\cite{Gysbers}. 
As one approaches $^{100}$Sn and the $Q_\mathrm{EC}$ values increase, all of the GT strength lies within the $Q_\mathrm{EC}$ window and the GT resonances can be studied in $\beta^+$ decay. Another aspect is that the GT strength centroids for even-even nuclei  are lower than for the odd-mass nuclei, and thus decay studies using beta-delayed high-resolution gamma-spectroscopy could be considered as reliable, while generally only total absorption gamma-spectroscopy can guarantee the correct estimation of all decay branches, especially the ones de-exciting by low-intensity high-energy gamma-cascades. This is known as the so-called Pandemonium problem (Refs. \cite{HARDY1977307,HARDY1984331}). For $^{100}$Sn ($N=Z=50$) so far the highest GT strengths have been reported \cite{GT, odd-even, Lubosss}  for the decay of $^{100}$Sn to the single state of $1^+$ in $^{100}$In. Figure~\ref{BGT} represents the GT strength of the even-even isotones on $N=50$ (top panel), where the main strength comes from a change of proton from the shell $1g_{9/2}$ to a neutron in shell $1g_{7/2}$ during the beta transition, separated by a large energy gap at $N=50$.\\
A new $Q_\mathrm{EC} = 5437\pm67$ for $^{98}$Cd nuclide was deduced from the mass of $^{98}$Cd measured in this work and the mass of daughter nuclide $^{98}$Ag from literature \cite{AME2020}. The new $Q_\mathrm{EC}$ value has an uncertainty that is less than half of that of the extrapolated value ($Q_\mathrm{EC} = 5330\pm140$) reported in Ref.~ \cite{98Cd-indirect}. The old values were obtained from the beta end-point energy measurements ($Q_\mathrm{EC} = 5430\pm40$) reported in a conference proceeding \cite{AIP-proceeding} and a non peer-reviewed GSI report \cite{98Cd-idirect-GSI}. Using the new $Q_\mathrm{EC}$ from this work, the improved half-life ($t_{1/2}=9.29\pm0.1$~s) from Ref.~\cite{PhysRevC.99.034313} and the log($ft$) values calculated from Ref.~\cite{logft} together with the recent decay scheme presented in Ref.~\cite{A98}, the summed Gamow-Teller strength ($B(\text{GT})$) for the five observed transitions ($0^+\longrightarrow1^+$) is $B(\text{GT})=2.94^{+0.32}_{-0.28}$ (red circle in figure~\ref{BGT}). The new GT strength has almost a factor 3 improved uncertainty and is still consistent with the previously reported value of $B(\text{GT})=3.5^{+0.8}_{-0.7}$ in Ref.~\cite{98Cd-indirect} and in agreement with the value reported in Refs. \cite{98Cd-idirect-GSI,AIP-proceeding} of $B(\text{GT})=2.9\pm0.2$. The new $B(\text{GT})$ value resulting from the direct mass measurement of $^{98}$Cd avoids possible systematic errors in beta end-point energy measurements of $Q$-values. In addition, we note that the previously reported $B(\text{GT})$ values were based only on observation of feeding to four transitions and a less-precise half-life~\cite{98Cd-indirect,98Cd-idirect-GSI,AIP-proceeding} (black star and black triangle at $Z=48$). 
The direct measurement reported here confirms the previous indirect and non peer-reviewed results, thus the general discussions with respect to the GT hindrance factor, in Refs. \cite{98Cd-indirect, Hu1999}  and most recently in Ref. \cite{Batist2010} are still correct.



The first attempt to describe the the Gamow-Teller strength in the vicinity of $^{100}$Sn was with the Shell Model Monte Carlo (SMMC) approach~\cite{DEAN199617} but provided only the total GT strength, and without assessing the spectroscopy and ph core excitations.  
State-of-the-art large scale shell (LSSM) model calculations within the $sdg$ valence space and GDSN interaction are presented in Figure~\ref{BGT}. The LSSM calculations allow configurations with up to 6p-6h excitations across the $Z=N=50$ closed shell, and  the ``standard'' quenching factor of $(0.75)^2 \approxeq 0.56$ as discussed above. These results are compared with the experimental values for the $N=50$ and $N=52$ (Figure~\ref{BGT}) isotones. In both, the protons and neutrons in the beta decay experience the same transition from $\pi 1g_{9/2}$ to $\nu 1g_{7/2}$, although for $N=52$ there is already a partial occupation of $\nu 1g_{7/2}$ shell which reduces the GT strength. Two sets of calculations are shown, either summing the GT strength for the observed  decaying states (solid line) or for the total strength contained in the Q$_{EC}$ window (dashed line), thus it indicates experimentally not observed GT strength. In both cases, the theoretical systematics suggest a relatively smooth trend towards and including $Z=50$. The approach to only considering the experimentally known transition in the $B(\text{GT})$ model calculations is done for the first time. It shows a remarkable good agreement between experiment and theory and clearly favors the $B(\text{GT})$ value as reported in the paper of Lubos et al., \cite{Lubosss}.

\section{Conclusions}

Mass measurements of isotopes near the $N=50$ shell closure have been performed, including 14 ground-state masses and two isomers. Among them, the mass of $^{98}$Cd and $^{97}$Rh and the excitation energy of $^{94m}$Rh have been measured directly. For $^{94}$Rh, four different shell model calculations have been performed and together with the measure excitation energy ($293\pm 21$~keV) allow to understand the level ordering  and spin-parity assignments of the ground and the isomeric states.
The controversy of the $Q_\mathrm{EC}$ values for $^{100}$Sn \cite{GT,Lubosss, nature-100In} was investigated with two approaches, the shifted two-neutron shell gap $\Delta_{2n} (Z,N+2)$ and the Gamow-Teller strength $B(\text{GT})$. The previously only indirectly measured isotopes that have an influence on $\Delta_{2n}(Z,N+2)$ at $N=50$ have been measured directly and confirm the previously known trends. So if one follows the arguments laid out in Ref. \cite{nature-100In}, this supports the $Q_\mathrm{EC}$ reported in Ref. \cite{GT}. 
In addition to this, a systematic comparison of experimental and theoretical $B(\text{GT})$ values for even-even isotones at $N=50$ and $N=52$ was performed.
For $^{98}$Cd the $B(\text{GT})$ value is determined to be $2.94^{+0.32}_{-0.28}$, for the first time by direct measurements. Large scale shell-model calculations have been performed for the full $B(\text{GT})$ strength and, more importantly, also only for the experimentally known transitions, enabling for the first time a direct comparison between theory and experiment. The experimental and theoretical value are in perfect agreement. The large scale shell-model calculations clearly support the $B(\text{GT})$ as reported in Ref. \cite{Lubosss}. The $\Delta_{2n}(Z,N+2)$ and $B(\text{GT})$ investigations for $^{100}$Sn come to contradictory results. Therefore, the current situation calls for new and improved experiments which could solve the ambiguity of the $Q_\mathrm{EC}$ of $^{100}$Sn and consequently $B$(GT) value, for example by a direct mass measurement of $^{100}$Sn at the FRS-IC.

\section*{Acknowledgments}
The authors acknowledge a long-standing collaboration with Hubert Grawe and John S. Winfield and deeply regret their passing. The results presented here are based on the experiment S474, which was performed at the FRS at the GSI Helmholtzzentrum f{\"u}r Schwerionenforschung, Darmstadt (Germany) in the context of FAIR Phase-0. The results presented in this paper are based on work performed before February 24$^{th}$ 2022.
This work was supported by the German Federal Ministry for Education and Research (BMBF) under contracts no.\ 05P19RGFN1 and 05P21RGFN1, by the German Research Foundation (DFG) under contract no.\ SCHE 1969/2-1, by the Hessian Ministry for Science and Art (HMWK) through the LOEWE Center HICforFAIR, by HGS-HIRe, and by Justus-Liebig-Universit{\"a}t Gie{\ss}en and GSI under the JLU-GSI strategic Helmholtzpartnership agreement. The paper was partly financed by the international project ``PMW'' of the Polish Minister of Science and Higher Education; active in the period 2022-2024; grant Nr 5237/GSI-FAIR/2022/0.





\bibliographystyle{apsrev4-1}
\bibliography{main.bib}

\begin{thebibliography}{74}%
\makeatletter
\providecommand \@ifxundefined [1]{%
 \@ifx{#1\undefined}
}%
\providecommand \@ifnum [1]{%
 \ifnum #1\expandafter \@firstoftwo
 \else \expandafter \@secondoftwo
 \fi
}%
\providecommand \@ifx [1]{%
 \ifx #1\expandafter \@firstoftwo
 \else \expandafter \@secondoftwo
 \fi
}%
\providecommand \natexlab [1]{#1}%
\providecommand \enquote  [1]{``#1''}%
\providecommand \bibnamefont  [1]{#1}%
\providecommand \bibfnamefont [1]{#1}%
\providecommand \citenamefont [1]{#1}%
\providecommand \href@noop [0]{\@secondoftwo}%
\providecommand \href [0]{\begingroup \@sanitize@url \@href}%
\providecommand \@href[1]{\@@startlink{#1}\@@href}%
\providecommand \@@href[1]{\endgroup#1\@@endlink}%
\providecommand \@sanitize@url [0]{\catcode `\\12\catcode `\$12\catcode
  `\&12\catcode `\#12\catcode `\^12\catcode `\_12\catcode `\%12\relax}%
\providecommand \@@startlink[1]{}%
\providecommand \@@endlink[0]{}%
\providecommand \url  [0]{\begingroup\@sanitize@url \@url }%
\providecommand \@url [1]{\endgroup\@href {#1}{\urlprefix }}%
\providecommand \urlprefix  [0]{URL }%
\providecommand \Eprint [0]{\href }%
\providecommand \doibase [0]{http://dx.doi.org/}%
\providecommand \selectlanguage [0]{\@gobble}%
\providecommand \bibinfo  [0]{\@secondoftwo}%
\providecommand \bibfield  [0]{\@secondoftwo}%
\providecommand \translation [1]{[#1]}%
\providecommand \BibitemOpen [0]{}%
\providecommand \bibitemStop [0]{}%
\providecommand \bibitemNoStop [0]{.\EOS\space}%
\providecommand \EOS [0]{\spacefactor3000\relax}%
\providecommand \BibitemShut  [1]{\csname bibitem#1\endcsname}%
\let\auto@bib@innerbib\@empty
\bibitem [{\citenamefont {Scheidenberger}(2005)}]{Scheidenberger2005}%
  \BibitemOpen
  \bibfield  {author} {\bibinfo {author} {\bibfnamefont {C.}~\bibnamefont
  {Scheidenberger}},\ }\href {\doibase 10.1016/j.nuclphysa.2005.02.006}
  {\bibfield  {journal} {\bibinfo  {journal} {Nuclear Physics A}\ }\textbf
  {\bibinfo {volume} {751}},\ \bibinfo {pages} {209} (\bibinfo {year}
  {2005})}\BibitemShut {NoStop}%
\bibitem [{\citenamefont {Isacker}\ \emph {et~al.}(1995)\citenamefont {Isacker}
  \emph {et~al.}}]{wigner}%
  \BibitemOpen
  \bibfield  {author} {\bibinfo {author} {\bibfnamefont {V.}~\bibnamefont
  {Isacker}} \emph {et~al.},\ }\href {\doibase 10.1103/PhysRevLett.74.4607}
  {\bibfield  {journal} {\bibinfo  {journal} {Phys. Rev. Lett.}\ }\textbf
  {\bibinfo {volume} {74}},\ \bibinfo {pages} {4607} (\bibinfo {year}
  {1995})}\BibitemShut {NoStop}%
\bibitem [{\citenamefont {Bentley}\ and\ \citenamefont
  {Frauendorf}(2013)}]{wigner2}%
  \BibitemOpen
  \bibfield  {author} {\bibinfo {author} {\bibfnamefont {I.}~\bibnamefont
  {Bentley}}\ and\ \bibinfo {author} {\bibfnamefont {S.}~\bibnamefont
  {Frauendorf}},\ }\href {\doibase 10.1103/PhysRevC.88.014322} {\bibfield
  {journal} {\bibinfo  {journal} {Physical Review C}\ }\textbf {\bibinfo
  {volume} {88}},\ \bibinfo {pages} {014322} (\bibinfo {year}
  {2013})}\BibitemShut {NoStop}%
\bibitem [{\citenamefont {Mukha}\ \emph {et~al.}(2006)\citenamefont {Mukha}
  \emph {et~al.}}]{94Ag-nature}%
  \BibitemOpen
  \bibfield  {author} {\bibinfo {author} {\bibfnamefont {I.}~\bibnamefont
  {Mukha}} \emph {et~al.},\ }\href {\doibase 10.1038/nature04453} {\bibfield
  {journal} {\bibinfo  {journal} {Nature}\ }\textbf {\bibinfo {volume} {439}},\
  \bibinfo {pages} {298} (\bibinfo {year} {2006})}\BibitemShut {NoStop}%
\bibitem [{\citenamefont {Hamaker}\ \emph {et~al.}(2021)\citenamefont {Hamaker}
  \emph {et~al.}}]{80Zr}%
  \BibitemOpen
  \bibfield  {author} {\bibinfo {author} {\bibfnamefont {A.}~\bibnamefont
  {Hamaker}} \emph {et~al.},\ }\href {\doibase 10.1038/s41567-021-01395-w}
  {\bibfield  {journal} {\bibinfo  {journal} {Nature Physics}\ }\textbf
  {\bibinfo {volume} {17}},\ \bibinfo {pages} {1408} (\bibinfo {year}
  {2021})}\BibitemShut {NoStop}%
\bibitem [{\citenamefont {Haettner}\ \emph {et~al.}(2011)\citenamefont
  {Haettner} \emph {et~al.}}]{Haettner2011b}%
  \BibitemOpen
  \bibfield  {author} {\bibinfo {author} {\bibfnamefont {E.}~\bibnamefont
  {Haettner}} \emph {et~al.},\ }\href {\doibase 10.1103/PhysRevLett.106.122501}
  {\bibfield  {journal} {\bibinfo  {journal} {{Phys. Rev. Lett.}}\ }\textbf
  {\bibinfo {volume} {{106}}},\ \bibinfo {pages} {{122501}} (\bibinfo {year}
  {{2011}})}\BibitemShut {NoStop}%
\bibitem [{\citenamefont {Faestermann}\ \emph {et~al.}(2013)\citenamefont
  {Faestermann} \emph {et~al.}}]{Faestermann2013}%
  \BibitemOpen
  \bibfield  {author} {\bibinfo {author} {\bibfnamefont {T.}~\bibnamefont
  {Faestermann}} \emph {et~al.},\ }\href {\doibase 10.1016/j.ppnp.2012.10.002}
  {\bibfield  {journal} {\bibinfo  {journal} {Prog. in Part. and Nucl. Phys.}\
  }\textbf {\bibinfo {volume} {69}},\ \bibinfo {pages} {85} (\bibinfo {year}
  {2013})}\BibitemShut {NoStop}%
\bibitem [{\citenamefont {Górska}(2022)}]{physics4010024}%
  \BibitemOpen
  \bibfield  {author} {\bibinfo {author} {\bibfnamefont {M.}~\bibnamefont
  {Górska}},\ }\href {\doibase 10.3390/physics4010024} {\bibfield  {journal}
  {\bibinfo  {journal} {Physics}\ }\textbf {\bibinfo {volume} {4}},\ \bibinfo
  {pages} {364} (\bibinfo {year} {2022})}\BibitemShut {NoStop}%
\bibitem [{\citenamefont {Hinke}\ \emph {et~al.}(2012)\citenamefont {Hinke}
  \emph {et~al.}}]{GT}%
  \BibitemOpen
  \bibfield  {author} {\bibinfo {author} {\bibfnamefont {C.~B.}\ \bibnamefont
  {Hinke}} \emph {et~al.},\ }\href {\doibase 10.1038/nature11116} {\bibfield
  {journal} {\bibinfo  {journal} {Nature}\ }\textbf {\bibinfo {volume} {486}},\
  \bibinfo {pages} {341} (\bibinfo {year} {2012})}\BibitemShut {NoStop}%
\bibitem [{\citenamefont {Lubos}\ \emph {et~al.}(2019)\citenamefont {Lubos}
  \emph {et~al.}}]{Lubosss}%
  \BibitemOpen
  \bibfield  {author} {\bibinfo {author} {\bibfnamefont {D.}~\bibnamefont
  {Lubos}} \emph {et~al.},\ }\href {\doibase 10.1103/PhysRevLett.122.222502}
  {\bibfield  {journal} {\bibinfo  {journal} {Phys. Rev. Lett.}\ }\textbf
  {\bibinfo {volume} {122}},\ \bibinfo {pages} {222502} (\bibinfo {year}
  {2019})}\BibitemShut {NoStop}%
\bibitem [{\citenamefont {Suzuki}\ \emph {et~al.}(2013)\citenamefont {Suzuki}
  \emph {et~al.}}]{18nbarn}%
  \BibitemOpen
  \bibfield  {author} {\bibinfo {author} {\bibfnamefont {H.}~\bibnamefont
  {Suzuki}} \emph {et~al.},\ }\href {\doibase
  https://doi.org/10.1016/j.nimb.2013.08.049} {\bibfield  {journal} {\bibinfo
  {journal} {Nucl. Instrum. Meth. B}\ }\textbf {\bibinfo {volume} {317}},\
  \bibinfo {pages} {756} (\bibinfo {year} {2013})}\BibitemShut {NoStop}%
\bibitem [{\citenamefont {Mougeot}\ \emph {et~al.}(2021)\citenamefont {Mougeot}
  \emph {et~al.}}]{nature-100In}%
  \BibitemOpen
  \bibfield  {author} {\bibinfo {author} {\bibfnamefont {M.}~\bibnamefont
  {Mougeot}} \emph {et~al.},\ }\href {\doibase 10.1038/s41567-021-01326-9}
  {\bibfield  {journal} {\bibinfo  {journal} {Nature Physics}\ }\textbf
  {\bibinfo {volume} {17}},\ \bibinfo {pages} {1099} (\bibinfo {year}
  {2021})}\BibitemShut {NoStop}%
\bibitem [{\citenamefont {Juodagalvis}\ \emph {et~al.}(2005)\citenamefont
  {Juodagalvis} \emph {et~al.}}]{odd-even}%
  \BibitemOpen
  \bibfield  {author} {\bibinfo {author} {\bibfnamefont {A.}~\bibnamefont
  {Juodagalvis}} \emph {et~al.},\ }\href {\doibase 10.1103/PhysRevC.72.024306}
  {\bibfield  {journal} {\bibinfo  {journal} {Physical Review C}\ }\textbf
  {\bibinfo {volume} {72}},\ \bibinfo {pages} {024306} (\bibinfo {year}
  {2005})}\BibitemShut {NoStop}%
\bibitem [{\citenamefont {Batist}\ \emph {et~al.}(2010)\citenamefont {Batist}
  \emph {et~al.}}]{Batist2010}%
  \BibitemOpen
  \bibfield  {author} {\bibinfo {author} {\bibfnamefont {L.}~\bibnamefont
  {Batist}} \emph {et~al.},\ }\href {\doibase 10.1140/epja/i2010-11025-x}
  {\bibfield  {journal} {\bibinfo  {journal} {The Eur. Phys. Jour. A}\ }\textbf
  {\bibinfo {volume} {46}},\ \bibinfo {pages} {45} (\bibinfo {year}
  {2010})}\BibitemShut {NoStop}%
\bibitem [{\citenamefont {Karny}\ \emph {et~al.}(1998)\citenamefont {Karny}
  \emph {et~al.}}]{KARNY19983}%
  \BibitemOpen
  \bibfield  {author} {\bibinfo {author} {\bibfnamefont {M.}~\bibnamefont
  {Karny}} \emph {et~al.},\ }\href {\doibase
  https://doi.org/10.1016/S0375-9474(98)00437-0} {\bibfield  {journal}
  {\bibinfo  {journal} {Nuclear Physics A}\ }\textbf {\bibinfo {volume}
  {640}},\ \bibinfo {pages} {3} (\bibinfo {year} {1998})}\BibitemShut {NoStop}%
\bibitem [{\citenamefont {Hu}\ \emph {et~al.}(1999{\natexlab{a}})\citenamefont
  {Hu} \emph {et~al.}}]{PhysRevC.60.024315}%
  \BibitemOpen
  \bibfield  {author} {\bibinfo {author} {\bibfnamefont {Z.}~\bibnamefont {Hu}}
  \emph {et~al.},\ }\href {\doibase 10.1103/PhysRevC.60.024315} {\bibfield
  {journal} {\bibinfo  {journal} {Phys. Rev. C}\ }\textbf {\bibinfo {volume}
  {60}},\ \bibinfo {pages} {024315} (\bibinfo {year}
  {1999}{\natexlab{a}})}\BibitemShut {NoStop}%
\bibitem [{\citenamefont {Walker}\ and\ \citenamefont
  {Podoly{\'a}k}(2020)}]{Walker2020}%
  \BibitemOpen
  \bibfield  {author} {\bibinfo {author} {\bibfnamefont {P.~M.}\ \bibnamefont
  {Walker}}\ and\ \bibinfo {author} {\bibfnamefont {Z.}~\bibnamefont
  {Podoly{\'a}k}},\ }\enquote {\bibinfo {title} {Nuclear isomers},}\ in\ \href
  {\doibase 10.1007/978-981-15-8818-1_46-1} {\emph {\bibinfo {booktitle}
  {Handbook of Nuclear Physics}}}\ (\bibinfo  {publisher} {Springer Nature
  Singapore},\ \bibinfo {address} {Singapore},\ \bibinfo {year}
  {2020})\BibitemShut {NoStop}%
\bibitem [{\citenamefont {Walker}(2010)}]{Walker2010}%
  \BibitemOpen
  \bibfield  {author} {\bibinfo {author} {\bibfnamefont {P.}~\bibnamefont
  {Walker}},\ }\href {\doibase 10.1016/j.nuclphysa.2009.12.017} {\bibfield
  {journal} {\bibinfo  {journal} {Nuclear Physics A}\ }\textbf {\bibinfo
  {volume} {834}},\ \bibinfo {pages} {22c} (\bibinfo {year}
  {2010})}\BibitemShut {NoStop}%
\bibitem [{\citenamefont {Dracoulis}\ \emph {et~al.}(2016)\citenamefont
  {Dracoulis} \emph {et~al.}}]{Dracoulis2016}%
  \BibitemOpen
  \bibfield  {author} {\bibinfo {author} {\bibfnamefont {G.~D.}\ \bibnamefont
  {Dracoulis}} \emph {et~al.},\ }\href {\doibase 10.1088/0034-4885/79/7/076301}
  {\bibfield  {journal} {\bibinfo  {journal} {Reports on Progress in Physics}\
  }\textbf {\bibinfo {volume} {79}},\ \bibinfo {pages} {076301} (\bibinfo
  {year} {2016})}\BibitemShut {NoStop}%
\bibitem [{\citenamefont {Chen}\ \emph {et~al.}(2013)\citenamefont {Chen} \emph
  {et~al.}}]{isomer_storage_ring}%
  \BibitemOpen
  \bibfield  {author} {\bibinfo {author} {\bibfnamefont {L.}~\bibnamefont
  {Chen}} \emph {et~al.},\ }\href {\doibase 10.1103/PhysRevLett.110.122502}
  {\bibfield  {journal} {\bibinfo  {journal} {Phys. Rev. Lett.}\ }\textbf
  {\bibinfo {volume} {110}},\ \bibinfo {pages} {122502} (\bibinfo {year}
  {2013})}\BibitemShut {NoStop}%
\bibitem [{\citenamefont {Franzke}\ \emph {et~al.}(2008)\citenamefont {Franzke}
  \emph {et~al.}}]{Franzke2008}%
  \BibitemOpen
  \bibfield  {author} {\bibinfo {author} {\bibfnamefont {B.}~\bibnamefont
  {Franzke}} \emph {et~al.},\ }\href {\doibase 10.1002/mas.20173} {\bibfield
  {journal} {\bibinfo  {journal} {Mass Spectrometry Reviews}\ }\textbf
  {\bibinfo {volume} {27}},\ \bibinfo {pages} {428} (\bibinfo {year}
  {2008})}\BibitemShut {NoStop}%
\bibitem [{\citenamefont {Block}\ \emph {et~al.}(2008)\citenamefont {Block}
  \emph {et~al.}}]{PhysRevLett.100.132501}%
  \BibitemOpen
  \bibfield  {author} {\bibinfo {author} {\bibfnamefont {M.}~\bibnamefont
  {Block}} \emph {et~al.},\ }\href {\doibase 10.1103/PhysRevLett.100.132501}
  {\bibfield  {journal} {\bibinfo  {journal} {Phys. Rev. Lett.}\ }\textbf
  {\bibinfo {volume} {100}},\ \bibinfo {pages} {132501} (\bibinfo {year}
  {2008})}\BibitemShut {NoStop}%
\bibitem [{\citenamefont {Blaum}(2006)}]{Blaum2006}%
  \BibitemOpen
  \bibfield  {author} {\bibinfo {author} {\bibfnamefont {K.}~\bibnamefont
  {Blaum}},\ }\href {\doibase 10.1016/j.physrep.2005.10.011} {\bibfield
  {journal} {\bibinfo  {journal} {Physics Reports}\ }\textbf {\bibinfo {volume}
  {425}},\ \bibinfo {pages} {1} (\bibinfo {year} {2006})}\BibitemShut {NoStop}%
\bibitem [{\citenamefont {Mardor}\ \emph {et~al.}(2021)\citenamefont {Mardor}
  \emph {et~al.}}]{Mardor-PRC}%
  \BibitemOpen
  \bibfield  {author} {\bibinfo {author} {\bibfnamefont {I.}~\bibnamefont
  {Mardor}} \emph {et~al.},\ }\href {\doibase 10.1103/PhysRevC.103.034319}
  {\bibfield  {journal} {\bibinfo  {journal} {Phys. Rev. C}\ }\textbf {\bibinfo
  {volume} {103}},\ \bibinfo {pages} {034319} (\bibinfo {year}
  {2021})}\BibitemShut {NoStop}%
\bibitem [{\citenamefont {Hornung}\ \emph {et~al.}(2020)\citenamefont {Hornung}
  \emph {et~al.}}]{Christine}%
  \BibitemOpen
  \bibfield  {author} {\bibinfo {author} {\bibfnamefont {C.}~\bibnamefont
  {Hornung}} \emph {et~al.},\ }\href {\doibase 10.1016/j.physletb.2020.135200}
  {\bibfield  {journal} {\bibinfo  {journal} {Physics Letters B}\ }\textbf
  {\bibinfo {volume} {802}},\ \bibinfo {pages} {135200} (\bibinfo {year}
  {2020})}\BibitemShut {NoStop}%
\bibitem [{\citenamefont {Beck}\ \emph {et~al.}(2021)\citenamefont {Beck} \emph
  {et~al.}}]{SoenkePRL}%
  \BibitemOpen
  \bibfield  {author} {\bibinfo {author} {\bibfnamefont {S.}~\bibnamefont
  {Beck}} \emph {et~al.},\ }\href {\doibase 10.1103/PhysRevLett.127.112501}
  {\bibfield  {journal} {\bibinfo  {journal} {Phys. Rev. Lett.}\ }\textbf
  {\bibinfo {volume} {127}},\ \bibinfo {pages} {112501} (\bibinfo {year}
  {2021})}\BibitemShut {NoStop}%
\bibitem [{\citenamefont {Geissel}\ \emph
  {et~al.}(1992{\natexlab{a}})\citenamefont {Geissel} \emph
  {et~al.}}]{Geissel1992}%
  \BibitemOpen
  \bibfield  {author} {\bibinfo {author} {\bibfnamefont {H.}~\bibnamefont
  {Geissel}} \emph {et~al.},\ }\href {\doibase 10.1103/PhysRevLett.68.3412}
  {\bibfield  {journal} {\bibinfo  {journal} {Phys. Rev. Lett.}\ }\textbf
  {\bibinfo {volume} {68}},\ \bibinfo {pages} {3412} (\bibinfo {year}
  {1992}{\natexlab{a}})}\BibitemShut {NoStop}%
\bibitem [{\citenamefont {Dickel}\ \emph {et~al.}(2015)\citenamefont {Dickel}
  \emph {et~al.}}]{Dickel2015b}%
  \BibitemOpen
  \bibfield  {author} {\bibinfo {author} {\bibfnamefont {T.}~\bibnamefont
  {Dickel}} \emph {et~al.},\ }\href {\doibase 10.1016/j.nima.2014.12.094}
  {\bibfield  {journal} {\bibinfo  {journal} {Nucl. Instrum. Meth. A}\ }\textbf
  {\bibinfo {volume} {{777}}},\ \bibinfo {pages} {172} (\bibinfo {year}
  {{2015}})}\BibitemShut {NoStop}%
\bibitem [{\citenamefont {Pla\ss{}}\ \emph {et~al.}(2008)\citenamefont
  {Pla\ss{}} \emph {et~al.}}]{Plass2008}%
  \BibitemOpen
  \bibfield  {author} {\bibinfo {author} {\bibfnamefont {W.~R.}\ \bibnamefont
  {Pla\ss{}}} \emph {et~al.},\ }\href {\doibase 10.1016/j.nimb.2008.05.079}
  {\bibfield  {journal} {\bibinfo  {journal} {Nucl. Instrum. Meth. B}\ }\textbf
  {\bibinfo {volume} {266}},\ \bibinfo {pages} {4560} (\bibinfo {year}
  {2008})}\BibitemShut {NoStop}%
\bibitem [{\citenamefont {Pla\ss{}}\ \emph {et~al.}(2013)\citenamefont
  {Pla\ss{}} \emph {et~al.}}]{Plass2013}%
  \BibitemOpen
  \bibfield  {author} {\bibinfo {author} {\bibfnamefont {W.~R.}\ \bibnamefont
  {Pla\ss{}}} \emph {et~al.},\ }\href {\doibase 10.1016/j.nimb.2013.07.063}
  {\bibfield  {journal} {\bibinfo  {journal} {Nucl. Instrum. Meth. B}\ }\textbf
  {\bibinfo {volume} {317}},\ \bibinfo {pages} {457} (\bibinfo {year}
  {2013})}\BibitemShut {NoStop}%
\bibitem [{\citenamefont {Geissel}\ \emph
  {et~al.}(1992{\natexlab{b}})\citenamefont {Geissel} \emph
  {et~al.}}]{Geissel1992b}%
  \BibitemOpen
  \bibfield  {author} {\bibinfo {author} {\bibfnamefont {H.}~\bibnamefont
  {Geissel}} \emph {et~al.},\ }\href {\doibase 10.1016/0168-583X(92)95944-M}
  {\bibfield  {journal} {\bibinfo  {journal} {Nucl. Instrum. Meth. B}\ }\textbf
  {\bibinfo {volume} {70}},\ \bibinfo {pages} {286 } (\bibinfo {year}
  {1992}{\natexlab{b}})}\BibitemShut {NoStop}%
\bibitem [{\citenamefont {Ranjan}\ \emph {et~al.}(2011)\citenamefont {Ranjan}
  \emph {et~al.}}]{Ranjan2011}%
  \BibitemOpen
  \bibfield  {author} {\bibinfo {author} {\bibfnamefont {M.}~\bibnamefont
  {Ranjan}} \emph {et~al.},\ }\href {\doibase 10.1209/0295-5075/96/52001}
  {\bibfield  {journal} {\bibinfo  {journal} {Europhys. Lett.}\ }\textbf
  {\bibinfo {volume} {96}},\ \bibinfo {pages} {52001} (\bibinfo {year}
  {2011})}\BibitemShut {NoStop}%
\bibitem [{\citenamefont {Purushothaman}\ \emph {et~al.}(2013)\citenamefont
  {Purushothaman} \emph {et~al.}}]{Purushothaman2013}%
  \BibitemOpen
  \bibfield  {author} {\bibinfo {author} {\bibfnamefont {S.}~\bibnamefont
  {Purushothaman}} \emph {et~al.},\ }\href {\doibase
  10.1209/0295-5075/104/42001} {\bibfield  {journal} {\bibinfo  {journal}
  {Europhys. Lett.}\ }\textbf {\bibinfo {volume} {104}},\ \bibinfo {pages}
  {42001} (\bibinfo {year} {2013})}\BibitemShut {NoStop}%
\bibitem [{\citenamefont {Ranjan}\ \emph {et~al.}(2015)\citenamefont {Ranjan}
  \emph {et~al.}}]{Ranjan2015}%
  \BibitemOpen
  \bibfield  {author} {\bibinfo {author} {\bibfnamefont {M.}~\bibnamefont
  {Ranjan}} \emph {et~al.},\ }\href {\doibase
  https://doi.org/10.1016/j.nima.2014.09.075} {\bibfield  {journal} {\bibinfo
  {journal} {Nucl. Instrum. Meth. A}\ }\textbf {\bibinfo {volume} {770}},\
  \bibinfo {pages} {87 } (\bibinfo {year} {2015})}\BibitemShut {NoStop}%
\bibitem [{\citenamefont {Reiter}(2015)}]{Reiter2015}%
  \BibitemOpen
  \bibfield  {author} {\bibinfo {author} {\bibfnamefont {M.~P.}\ \bibnamefont
  {Reiter}},\ }\emph {\bibinfo {title} {Pilot experiments with relativistic
  uranium projectile and fission fragments thermalized in a cryogenic
  gas-filled stopping cell}},\ \href
  {http://geb.uni-giessen.de/geb/volltexte/2015/11827} {Ph.D. thesis},\
  \bibinfo  {school} {Universit{\"a}t Gie\ss{}en} (\bibinfo {year}
  {2015})\BibitemShut {NoStop}%
\bibitem [{\citenamefont {Miskun}\ \emph {et~al.}(2015)\citenamefont {Miskun}
  \emph {et~al.}}]{Miskun2015}%
  \BibitemOpen
  \bibfield  {author} {\bibinfo {author} {\bibfnamefont {I.}~\bibnamefont
  {Miskun}} \emph {et~al.},\ }\href {\doibase
  10.15120/GR-2015-1-MU-NUSTAR-FRS-09} {\bibfield  {journal} {\bibinfo
  {journal} {GSI Sci. Rep. 2014}\ }\textbf {\bibinfo {volume} {2015-1}},\
  \bibinfo {pages} {146 p.} (\bibinfo {year} {2015})}\BibitemShut {NoStop}%
\bibitem [{\citenamefont {Haettner}\ \emph {et~al.}(2018)\citenamefont
  {Haettner} \emph {et~al.}}]{Haettner2018}%
  \BibitemOpen
  \bibfield  {author} {\bibinfo {author} {\bibfnamefont {E.}~\bibnamefont
  {Haettner}} \emph {et~al.},\ }\href {\doibase 10.1016/j.nima.2017.10.003}
  {\bibfield  {journal} {\bibinfo  {journal} {Nucl. Inst. Meth. A}\ }\textbf
  {\bibinfo {volume} {880}},\ \bibinfo {pages} {138} (\bibinfo {year}
  {2018})}\BibitemShut {NoStop}%
\bibitem [{\citenamefont {Ayet San~Andr\'es}\ \emph {et~al.}(2019)\citenamefont
  {Ayet San~Andr\'es} \emph {et~al.}}]{Ayet2019}%
  \BibitemOpen
  \bibfield  {author} {\bibinfo {author} {\bibfnamefont {S.}~\bibnamefont {Ayet
  San~Andr\'es}} \emph {et~al.},\ }\href {\doibase 10.1103/PhysRevC.99.064313}
  {\bibfield  {journal} {\bibinfo  {journal} {Phys. Rev. C}\ }\textbf {\bibinfo
  {volume} {99}},\ \bibinfo {pages} {064313} (\bibinfo {year}
  {2019})}\BibitemShut {NoStop}%
\bibitem [{\citenamefont {Purushothaman}\ \emph {et~al.}(2017)\citenamefont
  {Purushothaman} \emph {et~al.}}]{Purushothaman2017}%
  \BibitemOpen
  \bibfield  {author} {\bibinfo {author} {\bibfnamefont {S.}~\bibnamefont
  {Purushothaman}} \emph {et~al.},\ }\href {\doibase
  10.1016/j.ijms.2017.07.014} {\bibfield  {journal} {\bibinfo  {journal} {Int.
  J. Mass Spectrom.}\ }\textbf {\bibinfo {volume} {{421}}},\ \bibinfo {pages}
  {245} (\bibinfo {year} {{2017}})}\BibitemShut {NoStop}%
\bibitem [{\citenamefont {Kondev}\ \emph
  {et~al.}(2021{\natexlab{a}})\citenamefont {Kondev} \emph {et~al.}}]{AME2020}%
  \BibitemOpen
  \bibfield  {author} {\bibinfo {author} {\bibfnamefont {F.}~\bibnamefont
  {Kondev}} \emph {et~al.},\ }\href {\doibase 10.1088/1674-1137/abddae}
  {\bibfield  {journal} {\bibinfo  {journal} {Chinese Physics C}\ }\textbf
  {\bibinfo {volume} {45}},\ \bibinfo {pages} {030001} (\bibinfo {year}
  {2021}{\natexlab{a}})}\BibitemShut {NoStop}%
\bibitem [{\citenamefont {Kacker}\ \emph {et~al.}(2010)\citenamefont {Kacker}
  \emph {et~al.}}]{BR}%
  \BibitemOpen
  \bibfield  {author} {\bibinfo {author} {\bibfnamefont {R.~N.}\ \bibnamefont
  {Kacker}} \emph {et~al.},\ }\href {\doibase 10.6028/jres.115.031} {\bibfield
  {journal} {\bibinfo  {journal} {Nati. Inst. of Stan. and Tech.}\ }\textbf
  {\bibinfo {volume} {115}},\ \bibinfo {pages} {453} (\bibinfo {year}
  {2010})}\BibitemShut {NoStop}%
\bibitem [{\citenamefont {Kondev}\ \emph
  {et~al.}(2021{\natexlab{b}})\citenamefont {Kondev} \emph
  {et~al.}}]{NUBASE2020}%
  \BibitemOpen
  \bibfield  {author} {\bibinfo {author} {\bibfnamefont {F.}~\bibnamefont
  {Kondev}} \emph {et~al.},\ }\href {\doibase 10.1088/1674-1137/abddae}
  {\bibfield  {journal} {\bibinfo  {journal} {Chin. Phys. C}\ }\textbf
  {\bibinfo {volume} {45}},\ \bibinfo {pages} {030001} (\bibinfo {year}
  {2021}{\natexlab{b}})}\BibitemShut {NoStop}%
\bibitem [{\citenamefont {Oxorn}\ \emph {et~al.}(1980)\citenamefont {Oxorn}
  \emph {et~al.}}]{Oxorn1980}%
  \BibitemOpen
  \bibfield  {author} {\bibinfo {author} {\bibfnamefont {K.}~\bibnamefont
  {Oxorn}} \emph {et~al.},\ }\href {\doibase 10.1007/BF01434147} {\bibfield
  {journal} {\bibinfo  {journal} {Zeit. for Phys. A}\ }\textbf {\bibinfo
  {volume} {294}},\ \bibinfo {pages} {389} (\bibinfo {year}
  {1980})}\BibitemShut {NoStop}%
\bibitem [{\citenamefont {Herndl}\ \emph {et~al.}(1997)\citenamefont {Herndl}
  \emph {et~al.}}]{Herndl1997}%
  \BibitemOpen
  \bibfield  {author} {\bibinfo {author} {\bibfnamefont {H.}~\bibnamefont
  {Herndl}} \emph {et~al.},\ }\href {\doibase 10.1016/S0375-9474(97)00407-7}
  {\bibfield  {journal} {\bibinfo  {journal} {Nuclear Physics A}\ }\textbf
  {\bibinfo {volume} {627}},\ \bibinfo {pages} {35} (\bibinfo {year}
  {1997})}\BibitemShut {NoStop}%
\bibitem [{\citenamefont {Basu}\ \emph {et~al.}(1962)\citenamefont {Basu} \emph
  {et~al.}}]{97Rh-indirect1}%
  \BibitemOpen
  \bibfield  {author} {\bibinfo {author} {\bibfnamefont {B.}~\bibnamefont
  {Basu}} \emph {et~al.},\ }\href {\doibase 10.1016/0029-5582(62)90532-1}
  {\bibfield  {journal} {\bibinfo  {journal} {Nuclear Physics}\ }\textbf
  {\bibinfo {volume} {33}},\ \bibinfo {pages} {347 } (\bibinfo {year}
  {1962})}\BibitemShut {NoStop}%
\bibitem [{\citenamefont {Chikhladze}\ \emph {et~al.}(1962)\citenamefont
  {Chikhladze} \emph {et~al.}}]{97Rh-indirect2}%
  \BibitemOpen
  \bibfield  {author} {\bibinfo {author} {\bibfnamefont {V.~L.}\ \bibnamefont
  {Chikhladze}} \emph {et~al.},\ }\href@noop {} {\bibfield  {journal} {\bibinfo
   {journal} {Theoretical Physics}\ }\textbf {\bibinfo {volume} {43}},\
  \bibinfo {pages} {453} (\bibinfo {year} {1962})}\BibitemShut {NoStop}%
\bibitem [{\citenamefont {Plueckebaum}(1975)}]{97Rh-gamma}%
  \BibitemOpen
  \bibfield  {author} {\bibinfo {author} {\bibfnamefont {W.}~\bibnamefont
  {Plueckebaum}},\ }\href
  {https://inis.iaea.org/search/searchsinglerecord.aspx?recordsFor=SingleRecord&RN=7223132}
  {\bibfield  {journal} {\bibinfo  {journal} {Zeit. for Phys. A}\ }\textbf
  {\bibinfo {volume} {273}},\ \bibinfo {pages} {393} (\bibinfo {year}
  {1975})}\BibitemShut {NoStop}%
\bibitem [{\citenamefont {Stolz}\ \emph {et~al.}(2002)\citenamefont {Stolz}
  \emph {et~al.}}]{AIP-proceeding}%
  \BibitemOpen
  \bibfield  {author} {\bibinfo {author} {\bibfnamefont {A.}~\bibnamefont
  {Stolz}} \emph {et~al.},\ }\href {\doibase 10.1063/1.1517981} {\bibfield
  {journal} {\bibinfo  {journal} {AIP conference proceeding}\ }\textbf
  {\bibinfo {volume} {638}},\ \bibinfo {pages} {259} (\bibinfo {year}
  {2002})}\BibitemShut {NoStop}%
\bibitem [{\citenamefont {Stolz}\ \emph {et~al.}(2001)\citenamefont {Stolz}
  \emph {et~al.}}]{98Cd-idirect-GSI}%
  \BibitemOpen
  \bibfield  {author} {\bibinfo {author} {\bibfnamefont {A.}~\bibnamefont
  {Stolz}} \emph {et~al.},\ }\href {http://repository.gsi.de/record/53530}
  {\bibfield  {journal} {\bibinfo  {journal} {GSI Scientific Reports}\ }\textbf
  {\bibinfo {volume} {1}},\ \bibinfo {pages} {7} (\bibinfo {year}
  {2001})}\BibitemShut {NoStop}%
\bibitem [{\citenamefont {Plochocki}\ \emph {et~al.}(1992)\citenamefont
  {Plochocki} \emph {et~al.}}]{98Cd-indirect}%
  \BibitemOpen
  \bibfield  {author} {\bibinfo {author} {\bibfnamefont {A.}~\bibnamefont
  {Plochocki}} \emph {et~al.},\ }\href {\doibase 10.1007/BF01294487} {\bibfield
   {journal} {\bibinfo  {journal} {Zeit. for Phys. A}\ }\textbf {\bibinfo
  {volume} {342}},\ \bibinfo {pages} {43} (\bibinfo {year} {1992})}\BibitemShut
  {NoStop}%
\bibitem [{\citenamefont {{Ayet San Andr{\'e}s}}(2018)}]{Ayet2018}%
  \BibitemOpen
  \bibfield  {author} {\bibinfo {author} {\bibfnamefont {S.}~\bibnamefont
  {{Ayet San Andr{\'e}s}}},\ }\emph {\bibinfo {title} {{Developments for
  Multiple-Reflection Time-of-Flight Mass Spectrometers and their Application
  to High Resolution Mass Measurements of Exotic Nuclei}}},\ \href@noop {}
  {Ph.D. thesis},\ \bibinfo  {school} {Universit\"at Gie\ss{}en} (\bibinfo
  {year} {2018})\BibitemShut {NoStop}%
\bibitem [{\citenamefont {Weber}\ \emph {et~al.}(2008)\citenamefont {Weber}
  \emph {et~al.}}]{Weber2008}%
  \BibitemOpen
  \bibfield  {author} {\bibinfo {author} {\bibfnamefont {C.}~\bibnamefont
  {Weber}} \emph {et~al.},\ }\href {\doibase 10.1103/PhysRevC.78.054310}
  {\bibfield  {journal} {\bibinfo  {journal} {{Physical Review C}}\ }\textbf
  {\bibinfo {volume} {{78}}},\ \bibinfo {pages} {{054310}} (\bibinfo {year}
  {{2008}})}\BibitemShut {NoStop}%
\bibitem [{\citenamefont {Kurcewicz}\ \emph {et~al.}(1982)\citenamefont
  {Kurcewicz} \emph {et~al.}}]{Kurcewicz1982}%
  \BibitemOpen
  \bibfield  {author} {\bibinfo {author} {\bibfnamefont {W.}~\bibnamefont
  {Kurcewicz}} \emph {et~al.},\ }\href {\doibase 10.1007/BF01415845} {\bibfield
   {journal} {\bibinfo  {journal} {Zeit. for Phys. A}\ }\textbf {\bibinfo
  {volume} {308}},\ \bibinfo {pages} {21} (\bibinfo {year} {1982})}\BibitemShut
  {NoStop}%
\bibitem [{\citenamefont {Batist}\ \emph {et~al.}(2006)\citenamefont {Batist}
  \emph {et~al.}}]{Batist2006}%
  \BibitemOpen
  \bibfield  {author} {\bibinfo {author} {\bibfnamefont {L.}~\bibnamefont
  {Batist}} \emph {et~al.},\ }\href {\doibase 10.1140/epja/i2006-10074-0}
  {\bibfield  {journal} {\bibinfo  {journal} {{Eur. Phys. Jour. A}}\ }\textbf
  {\bibinfo {volume} {{29}}},\ \bibinfo {pages} {175} (\bibinfo {year}
  {{2006}})}\BibitemShut {NoStop}%
\bibitem [{\citenamefont {Clark}\ \emph {et~al.}(2005)\citenamefont {Clark}
  \emph {et~al.}}]{Clark2005}%
  \BibitemOpen
  \bibfield  {author} {\bibinfo {author} {\bibfnamefont {J.~A.}\ \bibnamefont
  {Clark}} \emph {et~al.},\ }\href {\doibase 10.1007/3-540-37642-9_177}
  {\bibfield  {journal} {\bibinfo  {journal} {{Eur. Phys. Jour. A}}\ }\textbf
  {\bibinfo {volume} {{25}}},\ \bibinfo {pages} {629} (\bibinfo {year}
  {{2005}})}\BibitemShut {NoStop}%
\bibitem [{\citenamefont {Fallis}\ \emph {et~al.}(2011)\citenamefont {Fallis}
  \emph {et~al.}}]{Fallis2011}%
  \BibitemOpen
  \bibfield  {author} {\bibinfo {author} {\bibfnamefont {J.}~\bibnamefont
  {Fallis}} \emph {et~al.},\ }\href {\doibase 10.1103/PhysRevC.84.045807}
  {\bibfield  {journal} {\bibinfo  {journal} {Phys. Rev. C}\ }\textbf {\bibinfo
  {volume} {84}},\ \bibinfo {pages} {045807} (\bibinfo {year}
  {2011})}\BibitemShut {NoStop}%
\bibitem [{\citenamefont {Abriola}\ \emph {et~al.}(2006)\citenamefont {Abriola}
  \emph {et~al.}}]{Abriola2006}%
  \BibitemOpen
  \bibfield  {author} {\bibinfo {author} {\bibfnamefont {D.}~\bibnamefont
  {Abriola}} \emph {et~al.},\ }\href {\doibase 10.1016/j.nds.2006.08.001}
  {\bibfield  {journal} {\bibinfo  {journal} {Nuclear Data Sheets}\ }\textbf
  {\bibinfo {volume} {107}},\ \bibinfo {pages} {2423} (\bibinfo {year}
  {2006})}\BibitemShut {NoStop}%
\bibitem [{\citenamefont {Gross}\ \emph {et~al.}(1976)\citenamefont {Gross}
  \emph {et~al.}}]{Gross1976}%
  \BibitemOpen
  \bibfield  {author} {\bibinfo {author} {\bibfnamefont {R.}~\bibnamefont
  {Gross}} \emph {et~al.},\ }\href {\doibase 10.1016/0375-9474(76)90645-X}
  {\bibfield  {journal} {\bibinfo  {journal} {Nuclear Physics A}\ }\textbf
  {\bibinfo {volume} {267}},\ \bibinfo {pages} {85} (\bibinfo {year}
  {1976})}\BibitemShut {NoStop}%
\bibitem [{\citenamefont {Honma}\ \emph {et~al.}(2009)\citenamefont {Honma}
  \emph {et~al.}}]{PhysRevC.80.064323}%
  \BibitemOpen
  \bibfield  {author} {\bibinfo {author} {\bibfnamefont {M.}~\bibnamefont
  {Honma}} \emph {et~al.},\ }\href {\doibase 10.1103/PhysRevC.80.064323}
  {\bibfield  {journal} {\bibinfo  {journal} {Phys. Rev. C}\ }\textbf {\bibinfo
  {volume} {80}},\ \bibinfo {pages} {064323} (\bibinfo {year}
  {2009})}\BibitemShut {NoStop}%
\bibitem [{\citenamefont {Blazhev}\ \emph {et~al.}(2004)\citenamefont {Blazhev}
  \emph {et~al.}}]{Blazhev:2004gm}%
  \BibitemOpen
  \bibfield  {author} {\bibinfo {author} {\bibfnamefont {A.}~\bibnamefont
  {Blazhev}} \emph {et~al.},\ }\href {\doibase 10.1103/PhysRevC.69.064304}
  {\bibfield  {journal} {\bibinfo  {journal} {Phys. Rev. C}\ }\textbf {\bibinfo
  {volume} {69}},\ \bibinfo {pages} {064304} (\bibinfo {year}
  {2004})}\BibitemShut {NoStop}%
\bibitem [{\citenamefont {Siciliano}\ \emph {et~al.}(2020)\citenamefont
  {Siciliano} \emph {et~al.}}]{Siciliano:2019qhw}%
  \BibitemOpen
  \bibfield  {author} {\bibinfo {author} {\bibfnamefont {M.}~\bibnamefont
  {Siciliano}} \emph {et~al.},\ }\href {\doibase
  10.1016/j.physletb.2020.135474} {\bibfield  {journal} {\bibinfo  {journal}
  {Phys. Lett. B}\ }\textbf {\bibinfo {volume} {806}},\ \bibinfo {pages}
  {135474} (\bibinfo {year} {2020})}\BibitemShut {NoStop}%
\bibitem [{\citenamefont {Grawe}\ \emph {et~al.}(2021)\citenamefont {Grawe}
  \emph {et~al.}}]{RISING:2021ait}%
  \BibitemOpen
  \bibfield  {author} {\bibinfo {author} {\bibfnamefont {H.}~\bibnamefont
  {Grawe}} \emph {et~al.} (\bibinfo {collaboration} {RISING}),\ }\href
  {\doibase 10.1016/j.physletb.2021.136591} {\bibfield  {journal} {\bibinfo
  {journal} {Phys. Lett. B}\ }\textbf {\bibinfo {volume} {820}},\ \bibinfo
  {pages} {136591} (\bibinfo {year} {2021})}\BibitemShut {NoStop}%
\bibitem [{\citenamefont {Towner}(1987)}]{TOWNER1987263}%
  \BibitemOpen
  \bibfield  {author} {\bibinfo {author} {\bibfnamefont {I.}~\bibnamefont
  {Towner}},\ }\href {\doibase 10.1016/0370-1573(87)90138-4} {\bibfield
  {journal} {\bibinfo  {journal} {Physics Reports}\ }\textbf {\bibinfo {volume}
  {155}},\ \bibinfo {pages} {263} (\bibinfo {year} {1987})}\BibitemShut
  {NoStop}%
\bibitem [{\citenamefont {Wildenthal}\ \emph {et~al.}(1983)\citenamefont
  {Wildenthal} \emph {et~al.}}]{PhysRevC.28.1343}%
  \BibitemOpen
  \bibfield  {author} {\bibinfo {author} {\bibfnamefont {B.~H.}\ \bibnamefont
  {Wildenthal}} \emph {et~al.},\ }\href {\doibase 10.1103/PhysRevC.28.1343}
  {\bibfield  {journal} {\bibinfo  {journal} {Phys. Rev. C}\ }\textbf {\bibinfo
  {volume} {28}},\ \bibinfo {pages} {1343} (\bibinfo {year}
  {1983})}\BibitemShut {NoStop}%
\bibitem [{\citenamefont {Chou}\ \emph {et~al.}(1993)\citenamefont {Chou} \emph
  {et~al.}}]{PhysRevC.47.163}%
  \BibitemOpen
  \bibfield  {author} {\bibinfo {author} {\bibfnamefont {W.-T.}\ \bibnamefont
  {Chou}} \emph {et~al.},\ }\href {\doibase 10.1103/PhysRevC.47.163} {\bibfield
   {journal} {\bibinfo  {journal} {Phys. Rev. C}\ }\textbf {\bibinfo {volume}
  {47}},\ \bibinfo {pages} {163} (\bibinfo {year} {1993})}\BibitemShut
  {NoStop}%
\bibitem [{\citenamefont {Mart\'{\i}nez-Pinedo}\ \emph
  {et~al.}(1996)\citenamefont {Mart\'{\i}nez-Pinedo} \emph
  {et~al.}}]{PhysRevC.53.R2602}%
  \BibitemOpen
  \bibfield  {author} {\bibinfo {author} {\bibfnamefont {G.}~\bibnamefont
  {Mart\'{\i}nez-Pinedo}} \emph {et~al.},\ }\href {\doibase
  10.1103/PhysRevC.53.R2602} {\bibfield  {journal} {\bibinfo  {journal} {Phys.
  Rev. C}\ }\textbf {\bibinfo {volume} {53}},\ \bibinfo {pages} {R2602}
  (\bibinfo {year} {1996})}\BibitemShut {NoStop}%
\bibitem [{\citenamefont {Gysbers}\ \emph {et~al.}(2019)\citenamefont {Gysbers}
  \emph {et~al.}}]{Gysbers}%
  \BibitemOpen
  \bibfield  {author} {\bibinfo {author} {\bibfnamefont {P.}~\bibnamefont
  {Gysbers}} \emph {et~al.},\ }\href {\doibase 10.1038/s41567-019-0450-7}
  {\bibfield  {journal} {\bibinfo  {journal} {Nature Physics}\ }\textbf
  {\bibinfo {volume} {15}},\ \bibinfo {pages} {428} (\bibinfo {year}
  {2019})}\BibitemShut {NoStop}%
\bibitem [{\citenamefont {Hardy}\ \emph {et~al.}(1977)\citenamefont {Hardy}
  \emph {et~al.}}]{HARDY1977307}%
  \BibitemOpen
  \bibfield  {author} {\bibinfo {author} {\bibfnamefont {J.}~\bibnamefont
  {Hardy}} \emph {et~al.},\ }\href {\doibase 10.1016/0370-2693(77)90223-4}
  {\bibfield  {journal} {\bibinfo  {journal} {Physics Letters B}\ }\textbf
  {\bibinfo {volume} {71}},\ \bibinfo {pages} {307} (\bibinfo {year}
  {1977})}\BibitemShut {NoStop}%
\bibitem [{\citenamefont {Hardy}\ \emph {et~al.}(1984)\citenamefont {Hardy}
  \emph {et~al.}}]{HARDY1984331}%
  \BibitemOpen
  \bibfield  {author} {\bibinfo {author} {\bibfnamefont {J.}~\bibnamefont
  {Hardy}} \emph {et~al.},\ }\href {\doibase 10.1016/0370-2693(84)92014-8}
  {\bibfield  {journal} {\bibinfo  {journal} {Physics Letters B}\ }\textbf
  {\bibinfo {volume} {136}},\ \bibinfo {pages} {331} (\bibinfo {year}
  {1984})}\BibitemShut {NoStop}%
\bibitem [{\citenamefont {Park}\ \emph {et~al.}(2019)\citenamefont {Park} \emph
  {et~al.}}]{PhysRevC.99.034313}%
  \BibitemOpen
  \bibfield  {author} {\bibinfo {author} {\bibfnamefont {J.}~\bibnamefont
  {Park}} \emph {et~al.},\ }\href {\doibase 10.1103/PhysRevC.99.034313}
  {\bibfield  {journal} {\bibinfo  {journal} {Phys. Rev. C}\ }\textbf {\bibinfo
  {volume} {99}},\ \bibinfo {pages} {034313} (\bibinfo {year}
  {2019})}\BibitemShut {NoStop}%
\bibitem [{log(nndc)}]{logft}%
  \BibitemOpen
  \href {https://www.nndc.bnl.gov/logft/} {\enquote {\bibinfo {title} {log(ft)
  calculator (https://www.nndc.bnl.gov/logft/)},}\ } (\bibinfo {year}
  {nndc})\BibitemShut {NoStop}%
\bibitem [{\citenamefont {Chen}\ \emph {et~al.}(2020)\citenamefont {Chen} \emph
  {et~al.}}]{A98}%
  \BibitemOpen
  \bibfield  {author} {\bibinfo {author} {\bibfnamefont {J.}~\bibnamefont
  {Chen}} \emph {et~al.},\ }\href {\doibase
  https://doi.org/10.1016/j.nds.2020.01.001} {\bibfield  {journal} {\bibinfo
  {journal} {Nuclear Data Sheets}\ }\textbf {\bibinfo {volume} {164}},\
  \bibinfo {pages} {1} (\bibinfo {year} {2020})}\BibitemShut {NoStop}%
\bibitem [{\citenamefont {Hu}\ \emph {et~al.}(1999{\natexlab{b}})\citenamefont
  {Hu} \emph {et~al.}}]{Hu1999}%
  \BibitemOpen
  \bibfield  {author} {\bibinfo {author} {\bibfnamefont {Z.}~\bibnamefont {Hu}}
  \emph {et~al.},\ }\href {\doibase 10.1103/PhysRevC.60.024315} {\bibfield
  {journal} {\bibinfo  {journal} {Physical Review C}\ }\textbf {\bibinfo
  {volume} {60}},\ \bibinfo {pages} {024315} (\bibinfo {year}
  {1999}{\natexlab{b}})}\BibitemShut {NoStop}%
\bibitem [{\citenamefont {Dean}\ \emph {et~al.}(1996)\citenamefont {Dean} \emph
  {et~al.}}]{DEAN199617}%
  \BibitemOpen
  \bibfield  {author} {\bibinfo {author} {\bibfnamefont {D.}~\bibnamefont
  {Dean}} \emph {et~al.},\ }\href {\doibase 10.1016/0370-2693(95)01446-2}
  {\bibfield  {journal} {\bibinfo  {journal} {Physics Letters B}\ }\textbf
  {\bibinfo {volume} {367}},\ \bibinfo {pages} {17} (\bibinfo {year}
  {1996})}\BibitemShut {NoStop}%
\end{thebibliography}%

\end{document}